\documentclass[journal]{IEEEtran}

\usepackage{algorithm}
\usepackage[noend]{algpseudocode}
\usepackage{array}
\usepackage[caption=false,font=normalsize,labelfont=sf,textfont=sf]{subfig}
\usepackage{textcomp}
\usepackage{stfloats}
\usepackage{url}
\usepackage{verbatim}
\usepackage{graphicx, xcolor}
\usepackage{cite}

\usepackage{bm}
\usepackage{amsmath,amssymb,amsfonts}
\usepackage{xcolor}
\usepackage{comment}
\usepackage{optidef}
\usepackage[T1]{fontenc}
\usepackage{mathtools} 
\usepackage{cuted}
\usepackage{flushend}
\usepackage{bbm}
\usepackage{dsfont}
\usepackage{mathtools}
\usepackage{multirow}
\usepackage{float}
\usepackage{dirtytalk}
\usepackage[hidelinks]{hyperref}
\usepackage{booktabs}
\usepackage{subcaption}
\usepackage{siunitx}
\usepackage{tcolorbox}

\usepackage{xcolor}

\tcbuselibrary{listings, skins, breakable}

\begin{document}

\bstctlcite{IEEEexample:BSTcontrol}

%\title{Interplay of Electromagnetic Interference and Inter-RIS Reflections in MISO Communications}
\title{On the Impact of Electromagnetic Interference and Inter-RIS Reflections in Indoor Factory Local 6G Networks}

\author{\IEEEauthorblockN{Ishan~Rangajith~Koralege, Nurul~Huda~Mahmood,~\IEEEmembership{Member, IEEE}, \\
Arthur~Sousa~de~Sena,~\IEEEmembership{ Member, IEEE}, and Italo~Atzeni,~\IEEEmembership{Senior Member, IEEE}}

\thanks{Ishan Rangajith Koralege, Nurul Huda Mahmood, and Italo Atzeni are with the Centre for Wireless Communications, University of Oulu, Finland. E-mail: \{ishan.koralege, nurulhuda.mahmood, italo.atzeni\}@oulu.fi.}% <-this % stops a space
\thanks{
Arthur Sousa de Sena is with Ericsson, 223 62 Lund, Sweden. E-mail: arthurssena@ieee.org.% <-this % stops a space
}
\thanks{The research leading to this paper was supported by the Research Council of Finland (336449 Profi6, 348396 HIGH-6G, 369116 \href{https://www.6gflagship.com/}{6G~Flagship}, and 359850 \href{https://nsf-6g-concorse.vt.domains/}{6G-ConCoRSe}) and Business Finland's 6GBridge program (8002/31/2022 Local~6G).}}
%\thanks{The research leading to this paper was supported by the Research Council of Finland (former Academy of Finland) \href{https://www.6gflagship.com/}{6G Flagship program} (Grant Number: 346208), and Business Finland's 6GBridge program through the projects Local 6G (Grant Number 8002/31/2022) and 6CORE (Grant Number 8410/31/2022).}}



\maketitle

\begin{abstract}
The sixth-generation (6G) radio technology is expected to include local networks as a special use case, extending the capabilities of `generic' 6G networks towards more demanding performance requirements. Reconfigurable intelligent surfaces (RISs) offer a novel paradigm for next-generation wireless communications, especially in the context of local 6G networks, enabling advanced signal propagation control through intelligent phase shift configurations. However, in practical deployments, their performance can be adversely affected by electromagnetic interference (EMI) from external sources and inter-RIS reflections (IRR) caused by signal reflections between multiple co-located RIS units. This paper presents a comprehensive analysis of the joint impact of EMI and IRR in a multi-RIS multi-cell system deployed within an indoor factory environment, while accounting for the spatial correlation among the associated channels. A detailed evaluation study is first carried out to investigate their impact on system performance. System-level simulations demonstrate that EMI is the dominant performance-limiting factor under moderate interference conditions. To address these adverse effects, an alternate optimization algorithm using the Riemannian conjugate gradient method is then proposed. The proposed algorithm optimizes the RIS phase shifts with and without interference information at the RIS, enabling a comparative evaluation of the resulting performance gains.
\end{abstract}


%%%%%%%%%%%%%%%%%%%%%%%%%%%%%%%
%%%%%%%%%%%%%%%%%%%%%%%%%%%%%%%
%%%                         %%%
%%%         SECTION         %%%
%%%                         %%%
%%%%%%%%%%%%%%%%%%%%%%%%%%%%%%%
%%%%%%%%%%%%%%%%%%%%%%%%%%%%%%%

\section{Introduction}
\label{sec:intro}

The sixth-generation (6G) radio technology is expected to support various novel characteristics in terms of operating spectrum, operational range, and served applications. A typical example is the \textit{local 6G network} paradigm, which operates in spatially constrained service areas such as an indoor factory setting. Reconfigurable intelligent surfaces (RIS) have gained significant attention as a potential technological solution~\cite{FarjamRIS, Koralege2025DeepUnfolding} to address the unique challenges of interference and radio resource management in local 6G scenarios~\cite{AMB+24_local6G}. An RIS comprises nearly passive reflecting elements capable of intelligently manipulating the wireless propagation environment~\cite{MBK23_functional}. Through adaptive control of the phase shifts of the RIS elements, the random wireless environment is passively transformed into a controllable channel, thereby rendering the signal propagation more deterministic. An RIS is particularly advantageous in local 6G networks like indoor factory environments, which often present complex propagation challenges~\cite{MBK23_functional, JuSubTerahertz2024}. However, many existing RIS studies tend to overlook practical system considerations, often leading to overly optimistic performance estimates. Therefore, evaluating RIS-enabled wireless systems under realistic deployment conditions in the context of local 6G networks (e.g., indoor factory environments) is essential to assess their true potential.

Electromagnetic interference (EMI) represents one of the primary challenges associated with practical RIS deployment~\cite{Emil2021electromagnetic, middleton2007statistical}. EMI comprises uncontrollable wireless signals originating from both human-made and natural sources. Since an RIS is unable to reflect signals selectively, ambient interference such as EMI is also reflected along with the desired transmitted signal. This is further exacerbated in indoor factory environments as the operation of heavy machinery and other industrial equipment generates EMI~\cite{StenumgaardMachineEMI}. Moreover, due to the close placement of RIS elements, spatial correlation among the associated channels becomes practically unavoidable~\cite{chandra2022downlink}. At the same time, to ensure comprehensive coverage and effectively support distributed users in complex indoor layouts, factories can deploy multiple RIS units close to each other. Due to this close arrangement, signals may reflect from one RIS unit to another before reaching a receiver, leading to inter-RIS reflections (IRR).

\subsection{Contributions}

In a densely obstructed environment, such as an indoor factory, the combined adverse effects of EMI and IRR must be considered to reflect a realistic scenario. Although a few previous studies have individually examined these phenomena in a general setting, to the best of our knowledge, no existing work has considered their coexisting impact. Motivated by this unexplored scenario, this study is the first to investigate the joint impact of EMI and IRR on the performance of an RIS-aided indoor industrial wireless network. Specifically, we perform a system-level analysis of their joint impact and thereafter propose a novel interference-aware algorithm that optimizes the RIS phase shifts to maximize the sum rate using the Riemannian conjugate gradient (RCG) method, effectively overcoming the detrimental impact of EMI and IRR. Our main contributions are summarized as follows:

\begin{itemize} 

\item We present a comprehensive analysis of the joint impact of EMI and IRR in multi-cluster, multi-user indoor factory environments aided by multiple RISs under various operating settings. We show, through asymptotic analysis and system-level evaluation, that EMI remains the dominant interference source at moderate transmit power levels, where IRR is negligible. However, at extremely high interference transmit powers or large neighboring RIS deployments, IRR becomes a consequential impairment that can no longer be overlooked. 

\item To overcome the detrimental impact of EMI and IRR in practical RIS deployment, we propose a novel RCG-based alternating optimization (AO) framework for RIS phase shift optimization in the presence of joint EMI and IRR. The proposed framework is found to provide sum rate gains up to 100\% and 400\% over the conventional projected gradient ascent (PGA) algorithm and the baseline unoptimized operation, respectively.

\item Unlike most existing works that assume uncorrelated channel matrices, we adopt a correlated channel model that better captures practical deployment conditions, where spatial correlation arises due to environmental and hardware constraints. The analysis and proposed optimization framework are extensively validated through numerical simulations across a wide range of system parameters, including the number of RIS elements, transmit power levels, RIS deployment configurations, EMI power levels, and IRR conditions.

\end{itemize}

The proposed interference-aware optimization framework effectively mitigates degradation and consistently outperforms interference-unaware optimization, especially as the EMI power and the number of RIS elements increase. Moreover, the performance gap between the interference-aware and interference-unaware frameworks becomes more pronounced as the Euclidean distance from the RIS to the base station (BS) and user equipments (UEs) decreases.

\subsection{Notations and Structure of the Paper}

Vectors are denoted by bold lower-case letters, while matrices are represented by bold upper-case letters. The Hermitian operator is denoted by $(\cdot)^\mathrm{H}$. The $i^{\text{th}}$ entry of a vector $\mathbf{x}$ is denoted by $[\mathbf{x}]_i$ and the $(i,j)^{\text{th}}$ entry of a matrix $\mathbf{X}$ by $[\mathbf{X}]_{ij}$. The Hadamard product is denoted by $\odot$, whereas the operator $\mathrm{diag}\{\cdot\}$ constructs a diagonal matrix from a given vector. The Euclidean norm is written as $\| \cdot \|_2$, while the magnitude of a scalar is denoted by $|\cdot|$. For optimization, the Riemannian gradient is expressed as $\mathrm{rgrad}\{\cdot\}$ and the standard gradient operator as $\nabla$. The expectation operator is denoted by $\mathbb{E}\{\cdot\}$ and the modulo operator by $\mathrm{mod}\{\cdot\}$. For a complex-valued quantity $x \in \mathbb{C}$, $\Re\{x\}$ and $\Im\{x\}$ denote its real and imaginary parts, respectively.

The rest of the paper is structured as follows. Section~\ref{sec:RelatedWorks} summarizes the related work in the literature. Section~\ref{sec:systemModel} introduces the system and channel models, together with the formulated expressions used in the simulation scenarios. The system-level analysis under different interference conditions including asymptotic analysis is presented in Section~\ref{sec:systemLevel}. To address those interferences, an AO framework is proposed in Section~\ref{sec:AlternateOptimization}. Based on this framework, comprehensive system-level simulations are carried out in Section~\ref{sec:PerfEvalOpt} to assess the overall performance. Finally, conclusions are drawn in Section~\ref{sec:conc}.

\section{Related Works}
\label{sec:RelatedWorks}

\subsection{EMI in RIS-Aided Systems}

The impact of EMI on the performance of RIS-aided systems has been investigated in a handful of works. The study in~\cite{Emil2021electromagnetic} introduced a heuristic RIS phase shift optimization aimed at jointly mitigating thermal noise and EMI, highlighting that EMI becomes increasingly critical as the number of RIS elements increases. Reference~\cite{khaleel2023electromagnetic} developed an EMI cancellation scheme for an RIS-assisted single antenna system, achieving superior performance over benchmark approaches. 
The study in~\cite{vega2022physical} explored the impact of EMI on the achievable secrecy performance of RIS-assisted wiretap systems under the influence of practical impairments. The findings showed that the influence of EMI on secrecy performance is highly dependent on the eavesdropper’s ability to suppress or cancel the interference. The work in~\cite{demir2024efficient} proposed an efficient channel estimation strategy for RIS-aided systems affected by EMI and spatial channel correlation, where the proposed optimized RIS configurations achieved significant improvements in channel estimation compared to conventional benchmarks. The work in \cite{Shi_RIS} provided a comprehensive overview of RIS-aided cell-free massive multiple-input multiple-output (MIMO) systems for 6G, highlighting key system design aspects and practical system considerations, including the impact of EMI in RIS-assisted wireless networks.

\subsection{IRR in RIS-Aided Systems}

The existing literature on IRR has revealed two prevailing perspectives concerning how these reflections are addressed within the system design. Most studies (e.g.,~\cite{liang2022multiRouteCascade, niu2021doubleRIS, han2022double, zhang2023double, ma2023multi}) have assumed multiple distributed RISs are deployed to assist in the desired signal transmission from a transmitter to one or more receivers. Hence, they have leveraged IRR as constructive signal propagation paths, enhancing overall system performance by improving signal strength, coverage, or spectral efficiency. In contrast, a smaller yet significant body of work (e.g.,~\cite{liu2023optimization, wang2023ris, nguyen2022leveraging}) has considered these reflections as sources of interference, emphasizing the need for dedicated mitigation or cancellation strategies. 

For example, the work in~\cite{nguyen2022leveraging} investigated a multi-user, multi-RIS uplink system that explicitly accounts for IRR, proposing an AO algorithm for joint phase shift and beamforming design. It further showed that, by effectively managing interference among closely deployed RISs, system throughput can be significantly improved, particularly when the number of RISs and reflecting elements is large.

\subsection{RIS Phase Shift Optimization}

RIS phase shift optimization is a widely studied topic. Usually, the phase shifts are optimized considering different objective functions, such as the sum rate, signal-to-interference-plus-
noise ratio (SINR), secrecy rate, energy efficiency, and others \cite{Peng_BeamformingOpt, Wang_hardImpOpt, Sui_StarOpt, Huang_EEOpt, Zhao_SecrecyOpt, Ginige2025MaxMin}. However, only a few of these works have incorporated EMI or IRR into the optimization framework, and none have addressed both simultaneously. In the study~\cite{liu2023optimization}, an AO approach was employed to jointly design the equalizer, transmit power, and RIS phase shift matrix in the presence of IRR. The work in~\cite{wang2023ris} also considered IRR and developed a power-efficient transmission strategy through AO of equalizers, transmit power, and RIS phase shifts. 

Reference~\cite{long2024mmse} designed the RIS phase shift coefficients to minimize the mean square error in both channel estimation and data transmission by leveraging knowledge of EMI, and proposed an AO framework based on the projected gradient method. Similarly, the study in~\cite{ma2023physical} employed an AO approach that jointly adjusts the beamforming vector and RIS phase shifts to eliminate EMI. In~\cite{kudumala2022hardware}, phase shift optimization was carried out in the presence of EMI to maximize the achievable sum rate using a particle swarm optimization algorithm.

\begin{figure}[t!]
    \centering
    \includegraphics[scale=0.33]{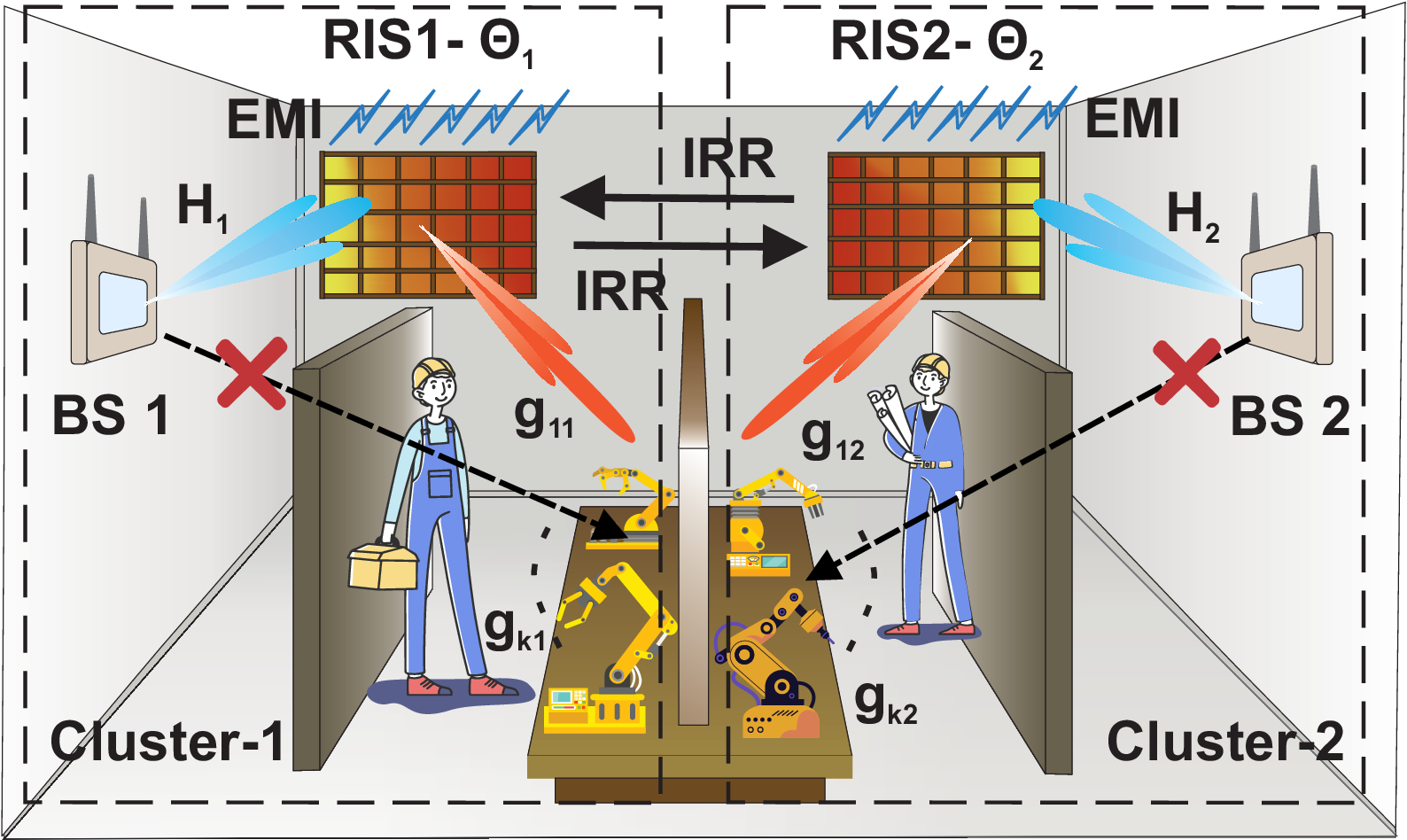}
    \caption{System model. An indoor factory environment assisted by multiple RISs.}
    \label{fig:system_model}
\end{figure}  

%%%%%%%%%%%%%%%%%%%%%%%%%%%%%%%
%%%%%%%%%%%%%%%%%%%%%%%%%%%%%%%
%%%                         %%%
%%%         SECTION         %%%
%%%                         %%%
%%%%%%%%%%%%%%%%%%%%%%%%%%%%%%%
%%%%%%%%%%%%%%%%%%%%%%%%%%%%%%%
\section{System Model and Formulations}
\label{sec:systemModel}

We consider a downlink multiple-input single-output (MISO) system in a quasi-static indoor factory environment, where multiple BSs serve multiple distributed UEs. The UEs are grouped into two clusters, denoted by \(n \in \{1,2\}\),\footnote{The methodologies and formulations developed in this study can be extended to more complex scenarios involving multiple clusters and additional RIS deployments. In this work, we restrict the analysis to two clusters to maintain clarity in the presentation and to enable meaningful insights into the joint behavior of EMI and IRR.} based on their spatial proximity and communication requirements. Due to dense obstacles in the factory, the direct BS-UE links are often blocked. To overcome this limitation and improve coverage, an RIS is strategically deployed in each cluster to enable reliable transmission, as illustrated in Fig.~\ref{fig:system_model}. Within the $n^{\text{th}}$ cluster, a BS equipped with $T_{n}$ antennas serves $K_n$ single-antenna users.

The transmitted signal from the BS in the $n^{\text{th}}$ cluster is defined as  
\begin{equation}
    \mathbf{x}_n[m] = \sum_{k=1}^{K_n} \mathbf{u}_{kn} \sqrt{p_{kn}} s_{kn}[m],
\end{equation}
 where \( \mathbf{u}_{kn} \in \mathbb{C}^{T_{n} \times 1} \) denotes the normalized beamforming vector satisfying \( \|\mathbf{u}_{kn} \| = 1 \), \( p_{kn} \) represents the transmit power allocated to the $k^{\text{th}}$ user in the $n^{\text{th}}$ cluster, and \( {s}_{kn}[m] \in \mathbb{C} \) is the normalized data symbol satisfying \( \mathbb{E}[|{s}_{kn}[m]|^2] = 1 \). The direct inter-cluster interference is assumed to be negligible due to signal blockage.  %In this configuration, we assume that significant blockages not only prevent interference from users in the external cluster but also ensure that no signals are received from external BSs. 

Let us assume that each RIS in the $n^{\text{th}}$ cluster consists of ${L}_n^2$ reflecting elements uniformly arranged in a ${L}_n \times {L}_n$ square grid and centered at the origin of its respective coordinate system. The elements of the $n^{\text{th}}$ RIS is assumed to have an area ${A}_n$, and the position of the $l^{\text{th}}$ element, indexed row-by-row, is denoted by the coordinate vector \(\mathbf{d}_{l}^{(n)} = [d_{x,l}^{(n)}, d_{y,l}^{(n)}, 0]^\mathrm{T}\), where the horizontal and vertical coordinates are given by \(d_{x,l}^{(n)} = -({L}_n - 1) \frac{\sqrt{A_n}}{2} + \sqrt{A_n} \mathrm{mod}(l - 1, {L}_n)\) and \(d_{y,l}^{(n)} = ({L}_n - 1) \frac{\sqrt{A_n}}{2} - \sqrt{A_n} \cdot \lfloor \frac{l - 1}{{L}_n} \rfloor\), respectively, with \(l \in \{1, 2, \ldots, {L}_n^2\}\). Each RIS element is capable of independently adjusting its reflection amplitude and phase shifts. The matrix of reflection coefficients of the $n^{\text{th}}$ RIS is expressed as
\begin{equation}
\boldsymbol{\Theta}_n = \mathrm{diag}(\eta_{n,1} e^{j \varphi_{n,1}}, \eta_{n,2} e^{j \varphi_{n,2}}, \ldots, \eta_{n,L_n^2} e^{j \varphi_{n,L_n^2}}),
\end{equation}
where \(\eta_{n,l} \in (0, 1]\) and \(\varphi_{n,l} \in [0, 2\pi)\) represent the amplitude and phase of the reflection coefficient associated with the $l^{\text{th}}$ element of the $n^{\text{th}}$ RIS, respectively. Assuming unit-modulus RIS elements, as commonly adopted in the literature (see, e.g., \cite{nguyen2022leveraging}), we focus exclusively on phase shift optimization to enable low-complexity and cost-effective implementations.

To characterize the path loss between all network components~\cite{3GPP_Standardized5G_Jiang_2021, StatisticalChannelInF_Ju}, we adopt the $3$GPP TR $38.901$ line-of-sight path loss model for indoor factory environments with sparse clutter and high base stations (InF-SH)~\cite{3gpp_tr38901}. Accordingly, the path loss coefficient for the link \( a \in \{\text{BS-UE}, \text{BS-RIS}, \text{RIS-UE}\} \) of the $n^{\text{th}}$ cluster is given by
\begin{equation}
\beta_a^{(n)}(d_{\text{3D}}) =31.84 
  + 21.50 \log_{10}(d_{\text{3D}}) 
  + 19.00 \log_{10}(f_{\text{c}}),
\label{eq:pl_los}
\end{equation}
where $d_{\text{3D}}$ denotes the three-dimensional distance in meters that incorporates BS and UE heights and $f_{\text{c}}$ represents the carrier frequency in GHz.

All the channel coefficients are subject to spatially correlated Rayleigh fading. The channel between the BS and the RIS associated with the $n^{\text{th}}$ cluster is denoted by \( \mathbf{H}_n \in \mathbb{C}^{L_n^2 \times T_n} \), while the channel vector between the RIS and the $k^{\text{th}}$ user in the $n^{\text{th}}$ cluster is represented by \( \mathbf{g}_{kn} = [g_{kn,1}, \ldots, g_{kn,L_n^2}]^T \in \mathbb{C}^{L_n^2 \times 1} \), where \( k \in \{1, 2, \ldots, K\} \). Both \( \mathbf{H}_n \) and \( \mathbf{g}_{kn} \) are modeled as spatially correlated channels, i.e., \(\operatorname{vec}(\mathbf{H}_n) \sim \mathcal{CN}\!(\mathbf{0},\; A_n\, \beta_{\mathrm{BS\text{-}RIS}}^{(n)}\, \mathbf{I}_{T_n} \otimes \mathbf{R}_{n})
\) and \( \mathbf{g}_{kn} \sim \mathcal{CN}(\mathbf{0}, A_n \beta_{\text{\tiny RIS-UE}}^{(n)} \mathbf{R}_n) \).  Here,  \( \mathbf{R}_n \) is the spatial correlation matrix associated with the $n^{\text{th}}$ RIS, with elements defined as~\cite{Emil2021electromagnetic}
\begin{equation}
[\mathbf{R}_n]_{l,\tilde{l}} = \mathrm{sinc}\left( \frac{2 \| \mathbf{d}_{l}^{(n)} - \mathbf{d}_{\tilde{l}}^{(n)} \|}{\lambda}\right),
\end{equation}
where $\mathbf{d}_{l}$ and $\mathbf{d}_{\tilde{l}} \in \mathbb{R}^3$ are the positions of the $l^{\text{th}}$ and ${\tilde{l}}^{\text{th}}$ elements, respectively, of the $n^{\text{th}}$ RIS and \( \lambda \) is the carrier wavelength. Next, we explore the scenarios influenced by the presence or absence of various types of external interference.

\subsection{Novel Joint EMI and IRR Scenario}
\label{subsec:EMI_IRR_RIS}

This subsection analyzes the joint impact of EMI and IRR by considering the complete system model shown in Fig.~\ref{fig:system_model}, a scenario that, to the best of our knowledge, has not been previously investigated. This comprehensive setup captures the simultaneous influence of EMI sources within the clusters and IRR between RIS~1 and RIS~2. As a result, the overall received signal under both EMI and IRR at the $k^{\text{th}}$ user in cluster~1 is given by
\begin{equation}\label{eq:received_signal_user_EMI_IRR1}
\begin{aligned}
y_{k1}^{\text{EMI+IRR}} = 
& \ \underbrace{
\mathbf{g}_{k1}^\mathrm{H} \boldsymbol{\Theta}_1 ( \mathbf{H}_{1} \mathbf{x}_1 + \mathbf{n}_1)
}_{\text{RIS reflection (via RIS 1)}} \\
&+ 
\underbrace{
\mathbf{g}_{k1}^\mathrm{H} \boldsymbol{\Theta}_1 
( \mathbf{Z}_{21}^\mathrm{H} \boldsymbol{\Theta}_2 ( \mathbf{H}_{2} \mathbf{x}_2 + \mathbf{n}_2 ))
}_{\text{IRR from RIS 2  reflected by RIS 1}} + n_{\text{w}},
\end{aligned}
\end{equation}
where $\mathbf{n}_n \sim \mathcal{CN}(\mathbf{0}, A_n \sigma_n^2 \mathbf{R}_n)$, \(n \in \{1,2\}\) denotes the EMI at the $n^{\text{th}}$ RIS, with covariance matrix $\mathbb{E}\{\mathbf{n}_n \mathbf{n}_n^\mathrm{H}\} = A_n \sigma_n^2 \mathbf{R}_n$, where $\sigma_n^2$ represents the corresponding EMI power. The additive white Gaussian noise (AWGN) is denoted by \( n_{\text{w}} \sim \mathcal{CN}(0, \sigma_{\text{w}}^2) \). The IRR from the $n^{\text{th}}$ RIS in cluster~\(n\) to the $m^{\text{th}}$ RIS in cluster~\(m\) is given by \( \operatorname{vec}( \mathbf{Z}_{nm} ) \sim \mathcal{CN}( \mathbf{0}, \sqrt {A_nA_m} \beta_{nm} \mathbf{I}_{L_1^2 L_2^2}) \). Here, \( \beta_{nm} \) denotes path loss coefficient between the $n^{\text{th}}$ RIS and the $m^{\text{th}}$ RIS, following the same model as in \eqref{eq:pl_los}. The identity matrix \( \mathbf{I}_{L_1^2 L_2^2} \) is used under the assumption that spatial correlation between RISs is negligible, as they are designed to serve distinct and sufficiently separated clusters. Moreover, \eqref{eq:received_signal_user_EMI_IRR1} can be further expanded as in \eqref{eq:received_signal_user_EMI_IRR2} at the top of the page, resulting in the SINR of the $k^{\text{th}}$ expressed as in \eqref{eq:sinr_expression_EMI_IRR} at the top of the page. The above formulation provides a unified system model. In the subsequent subsections, it is specialized to distinct scenarios, namely the EMI-only, IRR-only, and external-interference-free (EIF) cases.

\begin{figure*}[t!]
\begin{equation}\label{eq:received_signal_user_EMI_IRR2}
\begin{aligned}
y_{k1}^{\text{EMI+IRR}} = 
& \ \underbrace{
\mathbf{g_{k1}}^\mathrm{H} \boldsymbol{\Theta_1} \mathbf{H_{1}}\mathbf{u_{1k}} \sqrt{p_{k1}} d_{k1}
}_{\text{Desired signal (via RIS 1)}}
+ 
\underbrace{
\mathbf{g_{k1}}^\mathrm{H} \boldsymbol{\Theta_1} \mathbf{H_{1}}\sum_{i \neq k} \mathbf{u_{i1}} \sqrt{p_{i1}} d_{i1}
}_{\text{Intra-cluster interference}}
+ 
\underbrace{
\mathbf{g_{k1}}^\mathrm{H} \boldsymbol{\Theta_1} \mathbf{Z_{21}}^\mathrm{H} \boldsymbol{\Theta_2} \mathbf{H_{2}} \sum_{j=1}^{N} \mathbf{u_{j2}} \sqrt{p_{j2}} d_{j2}
}_{\text{IRR via RIS 2 and RIS 1}}
\\
&+ 
\underbrace{
\mathbf{g_{k1}}^\mathrm{H} \boldsymbol{\Theta_1} \mathbf{n_1}
}_{\text{Noise reflected by RIS 1}}
+ 
\underbrace{
\mathbf{g_{k1}}^\mathrm{H} \boldsymbol{\Theta_1} \mathbf{Z_{21}}^\mathrm{H} \boldsymbol{\Theta_2} \mathbf{n_2}
}_{\text{Noise from cluster~2 reflected via RIS~2 and RIS~1}}
+ 
n_{\text{w}}
.
\end{aligned}
\end{equation}

\vspace{0.5em} % small vertical space between equations

\begin{equation}\label{eq:sinr_expression_EMI_IRR}
\begin{split}
\gamma_{k1}^{\text{EMI+IRR}} = & \ \frac{
p_{k1} | \mathbf{g_{k1}}^\mathrm{H} \boldsymbol{\Theta_1} \mathbf{H_{1}} \mathbf{u_{k1}}|^2
}{
\sum_{i \neq k} p_{i1} | \mathbf{g_{k1}}^\mathrm{H} \boldsymbol{\Theta_1} \mathbf{H_{1}} \mathbf{u_{i1}} |^2
+ 
\sum_{j=1}^{N} p_{j2} | \mathbf{g_{k1}}^\mathrm{H} \boldsymbol{\Theta_1} \mathbf{Z_{21}}^\mathrm{H} \boldsymbol{\Theta_2} \mathbf{H_{2}} \mathbf{u_{j2}} |^2
+
\mathbf{g_{k1}}^\mathrm{H} \boldsymbol{\Theta_1} A_1 \sigma_1^2 \mathbf{R_1} \boldsymbol{\Theta_1}^\mathrm{H} \mathbf{g_{k1}}
 + \ldots }\\
&\hspace{10em} \mathbf{g_{k1}}^\mathrm{H} \boldsymbol{\Theta_1} \mathbf{Z_{21}}^\mathrm{H} \boldsymbol{\Theta_2} A_2 \sigma_2^2 \mathbf{R_2} \boldsymbol{\Theta_2}^\mathrm{H} \mathbf{Z_{21}} \boldsymbol{\Theta_1}^\mathrm{H} \mathbf{g_{k1}}
+ \sigma_{\text{w}}^2.
\end{split}
\end{equation}
\hrulefill
\end{figure*}

\subsection{IRR-Only Scenario}
\label{subsec:IRR_RIS}
The above analysis is now extended to a scenario where external EMI sources are absent. In this case, the effects of EMI within both clusters are neglected to isolate and examine the system behavior solely in the presence of IRR. Accordingly, we set $\mathbf{n}_n = \mathbf{0}$, \(n \in \{1,2\}\) in \eqref{eq:received_signal_user_EMI_IRR1}. Under these assumptions, the overall received signal at the $k^{\text{th}}$ user in cluster~1 is given by \eqref {eq:received_signal_user_IRR2}, leading to the corresponding SINR expressed as in \eqref{eq:sinr_expression_IRR} at the top of the next page. We next analyze cluster~1 in isolation by assuming that cluster~2 is absent. Under this assumption, the effect of IRR is eliminated.

\subsection{EMI-Only Scenario}
\label{subsec:EMI_RIS}

\setcounter{equation}{9}
In this scenario, external EMI sources in cluster~1 are assumed to be present, thereby simplifying the analysis and enabling a clearer assessment of system behavior under the sole influence of EMI. Accordingly, we set $\mathbf{n}_2 = \mathbf{0}$ and $\mathbf{Z}_{21} = \mathbf{0}$ in \eqref{eq:received_signal_user_EMI_IRR1}. Under these conditions, the received signal at the $k^{\text{th}}$ user in cluster~1 in the presence of EMI is given by
\begin{equation}\label{eq:received_signal_user_EMI}
\begin{aligned}
y_{k1}^{\text{EMI}} = 
& \ \underbrace{
\mathbf{g}_{k1}^\mathrm{H} \boldsymbol{\Theta}_1 \mathbf{H}_{1}
\mathbf{u}_{k1} \sqrt{p_{k1}} d_{k1}
}_{\text{Desired signal (via RIS 1)}}
+
\underbrace{
\mathbf{g}_{k1}^\mathrm{H} \boldsymbol{\Theta}_1 \mathbf{H}_{1}
\sum_{i \neq k} \mathbf{u}_{i1} \sqrt{p_{i1}} d_{i1}
}_{\text{Intra-cluster interference}}
\\
&+
\underbrace{
\mathbf{g}_{k1}^\mathrm{H} \boldsymbol{\Theta}_1 \mathbf{n}_1
}_{\text{Noise reflected by RIS 1}}
+ n_{\text{w}}.
\end{aligned}
\end{equation}
Using \eqref{eq:received_signal_user_EMI}, the SINR for the $k^{\text{th}}$ user can be derived as in \eqref{eq:sinr_expression_EMI} at the top of the next page. 

\subsection{External-Interference-Free Scenario}
\label{subsec:IF}
\setcounter{equation}{11}
In the absence of EMI and IRR, the system reduces to a scenario where the corresponding interference terms are zero, which can be achieved by setting $\mathbf{n}_1 = \mathbf{n}_2 = \mathbf{0}$ and $\mathbf{Z}_{21} = \mathbf{0}$ in \eqref{eq:received_signal_user_EMI_IRR1} Under these assumptions, the signal received by the $k^{\text{th}}$ user in cluster~1 is given by
\begin{equation}\label{eq:received_signal_user_IF}
\begin{aligned}
y_{k1}^{\text{EIF}} = 
& \ \underbrace{
\mathbf{g}_{k1}^\mathrm{H} \boldsymbol{\Theta}_1 \mathbf{H}_{1}
\mathbf{u}_{k1} \sqrt{p_{k1}} d_{k1}
}_{\text{Desired signal (via RIS 1)}}
+
\underbrace{
\mathbf{g}_{k1}^\mathrm{H} \boldsymbol{\Theta}_1 \mathbf{H}_{1}
\sum_{i \neq k} \mathbf{u}_{i1} \sqrt{p_{i1}} d_{i1}
}_{\text{Intra-cluster interference}}
\\
&+ n_{\text{w}}.
\end{aligned}
\end{equation}
Then the SINR under the EIF scenario for the $k^{\text{th}}$ user is given by
\begin{equation} \label{eq:sinr_expression}
\gamma_{k1}^{\text{EIF}} = \frac{p_{k1} | \mathbf{g}_{k1}^\mathrm{H} \boldsymbol{\Theta}_1 \mathbf{H}_1 \mathbf{u}_{k1} |^2}
{\sum_{i \neq k} p_{i1} |\mathbf{g}_{k1}^\mathrm{H} \boldsymbol{\Theta}_1 \mathbf{H}_1\mathbf{u}_{i1}|^2 + \sigma_{\text{w}}^2}.
\end{equation}

Deriving analytical performance trends for the considered system model would require overly simplifying assumptions to ensure analytical tractability, yielding results that are trivial and offer little practical insight. We therefore present system-level performance evaluation, as detailed in the following section, consistent with the methodology adopted in~\cite{Zhang_massiveMIMO, Dai_twotimescale}.

%%%%%%%%%%%%%%%%%%%%%%%%%%%%%%%
%%%%%%%%%%%%%%%%%%%%%%%%%%%%%%%
%%%                         %%%
%%%         SECTION         %%%
%%%                         %%%
%%%%%%%%%%%%%%%%%%%%%%%%%%%%%%%
%%%%%%%%%%%%%%%%%%%%%%%%%%%%%%%
\section{System-Level Evaluation}
\label{sec:systemLevel}

\begin{figure*}[t!]

\setcounter{equation}{7}

\begin{equation}\label{eq:received_signal_user_IRR2}
y_{k1}^{\text{IRR}} =\; 
\underbrace{\mathbf{g}_{k1}^\mathrm{H} \boldsymbol{\Theta}_1 \mathbf{H}_{1} \mathbf{u}_{k1} \sqrt{p_{k1}} d_{k1}}_{\text{Desired signal (via RIS 1)}}
+ 
\underbrace{\mathbf{g}_{k1}^\mathrm{H} \boldsymbol{\Theta}_1 \mathbf{H}_{1} \sum_{i \neq k} \mathbf{u}_{i1} \sqrt{p_{i1}} d_{i1}}_{\text{Intra-cluster interference}}
+ 
\underbrace{\mathbf{g}_{k1}^\mathrm{H} \boldsymbol{\Theta}_1 \mathbf{Z}_{21}^\mathrm{H} \boldsymbol{\Theta}_2 \mathbf{H}_{2} \sum_{j=1}^{N} \mathbf{u}_{j2} \sqrt{p_{j2}} d_{j2}}_{\text{IRR via RIS 2 and RIS 1}}
+ n_{\text{w}}.
\end{equation}

\vspace{0.5em} % small vertical space between equations

\begin{equation}\label{eq:sinr_expression_IRR}
\begin{aligned}
\gamma_{k1}^{\text{IRR}} =\; 
\frac{
p_{k1} | \mathbf{g}_{k1}^\mathrm{H} \boldsymbol{\Theta}_1 \mathbf{H}_{1} \mathbf{u}_{k1}|^2
}{
\sum_{i \neq k} p_{i1} | \mathbf{g}_{k1}^\mathrm{H} \boldsymbol{\Theta}_1 \mathbf{H}_{1} \mathbf{u}_{i1} |^2 
+ 
\sum_{j=1}^{N} p_{j2} | \mathbf{g}_{k1}^\mathrm{H} \boldsymbol{\Theta}_1 \mathbf{Z}_{21}^\mathrm{H} \boldsymbol{\Theta}_2 \mathbf{H}_{2} \mathbf{u}_{j2} |^2 
+ \sigma_{\text{w}}^2
}.
\end{aligned}
\end{equation}

\vspace{0.5em} % small vertical space between equations
\setcounter{equation}{10}
\begin{equation} \label{eq:sinr_expression_EMI}
\gamma_{k1}^{\text{EMI}} = \frac{p_{k1} | \mathbf{g}_{k1}^\mathrm{H} \boldsymbol{\Theta}_1 \mathbf{H}_1 \mathbf{u}_{k1}|^2}
{\sum_{i \neq k} p_{i1} |\mathbf{g}_{k1}^\mathrm{H} \boldsymbol{\Theta}_1 \mathbf{H}_1 \mathbf{u}_{i1} |^2 
+ \mathbf{g}_{k1}^\mathrm{H} \boldsymbol{\Theta}_1{A}_1 \sigma_1^2 \mathbf{R}_1 \boldsymbol{\Theta}_1^\mathrm{H} \mathbf{g}_{k1} + \sigma_{\text{w}}^2}.
\end{equation}

\hrulefill
\end{figure*}

To investigate the impact of EMI and IRR on system performance, we first conduct a numerical evaluation using the system setup illustrated in Fig.~\ref{fig:system_model}. In particular, the effects of EMI, IRR, and their joint presence are examined. To complement these evaluations and extract tractable design insights, an asymptotic analysis is subsequently developed, characterizing the limiting SINR behavior as the number of RIS elements grows large. This analysis establishes the necessary insights and motivation prior to presenting the proposed RCG-based AO algorithm, which represents one of the main contributions of this work. 
In our performance evaluation, we vary the parameters within cluster~1 while keeping all parameters of cluster~2 fixed. We consider a system operating with a thermal noise power spectral density of $N_0 = -174$~dBm/Hz. Consequently, the noise power is calculated as $\sigma_{\text{w}}^2 = N_0 B = -114$~dBm. The reference area of a reflecting element is $A_1 = A_2 = \big(\frac{\lambda}{4}\big)^2 = 6.25~\times~ 10^{-4}~$m\textsuperscript{2}.
The phase shifts of both RISs are fixed to zero. For the scenarios considering EMI, we consider two EMI levels: $A_n\sigma_n^2 = -75$~dBm and  $A_n\sigma_n^2 =-65$~dBm. The EMI is assumed to be isotropic. 

\begin{figure}[t!]
    \centering
    \includegraphics[scale=0.48]{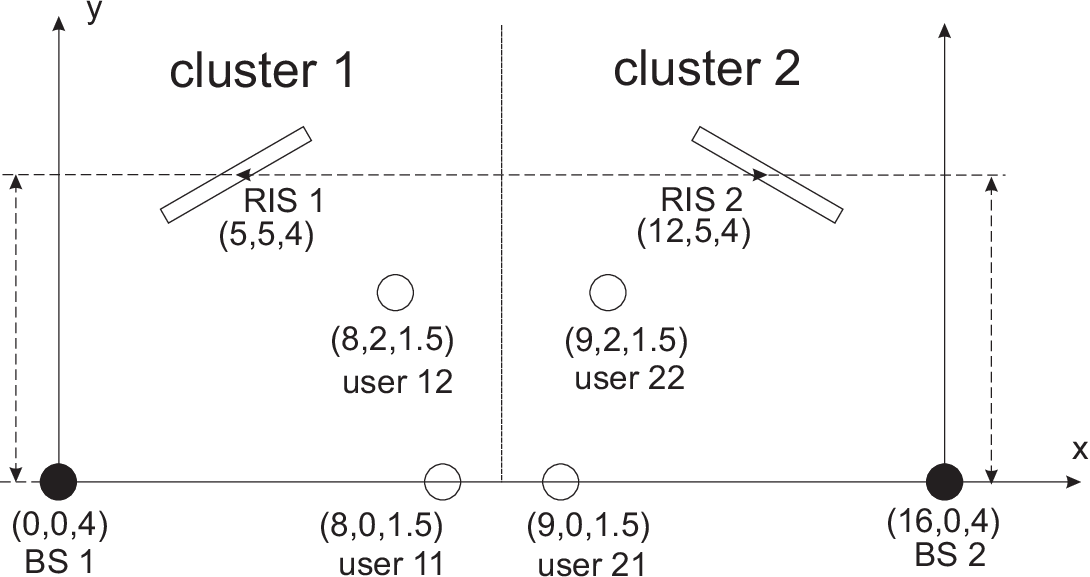}
    \caption{Simulation coordinate system.}
    \label{fig:Model_cordinates}
\end{figure}

Although industrial EMI is often non-stationary and impulsive, we adopt a stationary Gaussian model to obtain a tractable baseline and conservative lower-bound for performance evaluation. This simplified indoor factory model serves to isolate and analyze the joint impact of EMI and IRR under controlled conditions. While more detailed interference models may yield different performance gains, the main trends and insights are expected to remain the same. The coordinates used in the system-level evaluations are illustrated in Fig.~\ref{fig:Model_cordinates} and other simulation parameters are summarized in Table~\ref{tab:parameters}.

\subsection{Numerical Results} 
\label{SimAnalysis}

\begin{table}[t!]
\centering
\caption{Simulation parameters.}
\begin{tabular}{|l|l|l|}
\hline
\textbf{Parameter} & \textbf{Description} & \textbf{Value} \\
\hline
$f_{c1}, f_{c2}$ & Carrier frequency & 3~GHz \\
\hline
$\lambda_{1}, \lambda_{2}$ & Wavelength & 0.1~m \\
\hline
$B$ & Bandwidth & 1~MHz \\
\hline
$P_{2}$ & Transmit power of BS~2 & 30~dBm \\
\hline
$L_{2}^2$ & Elements of RIS 2 & 400 \\
\hline
$h_{\textrm{BS} 1}, h_{\textrm{BS} 2}$ & Height of BSs & 4~m \\
\hline
$h_{\textrm{RIS} 1}, h_{\textrm{RIS} 2}$ & Height of RISs & 4~m \\
\hline
$h_{\textrm{UEs}}$ & Height of UEs & 1.5~m \\
\hline
\end{tabular}
\label{tab:parameters}
\end{table}

\begin{figure}[t!]
    \centering    \includegraphics[width=\columnwidth]{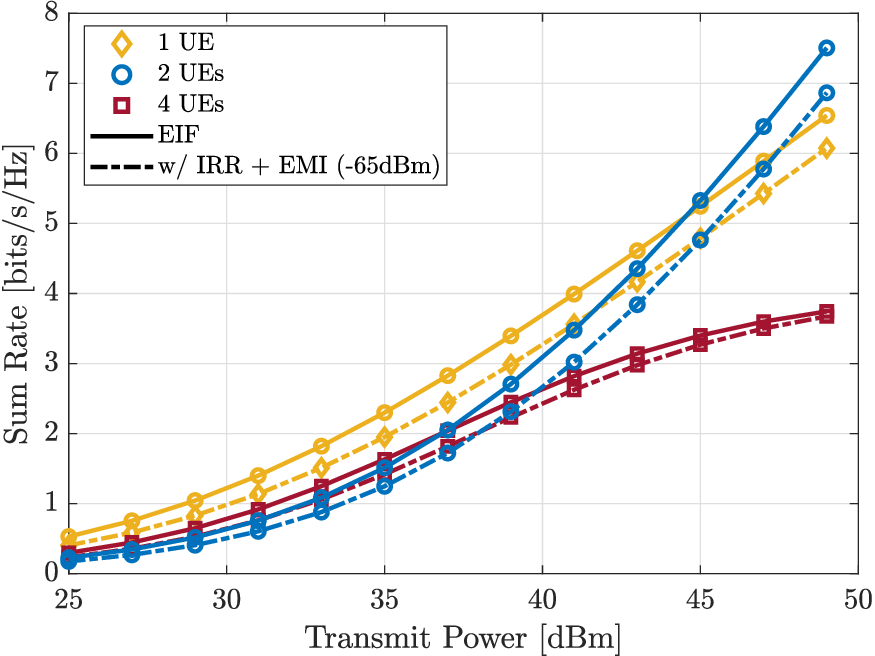}
    %\captionsetup{font=tiny}
    \caption{Sum rate vs transmit power for different number of UEs.}
    \label{fig:Result3}
\end{figure}

Considering the dense deployment of UEs in an indoor factory environment, different numbers of UEs are examined. The average rate is evaluated via Monte Carlo simulation over $3 \times 10^3$ independent channel realizations. The channels $\mathbf{H}_1$, $\mathbf{H}_2$, $\mathbf{g}_{k1}$, and $\mathbf{g}_{k2}$ are modeled as spatially isotropic, with covariance matrices $R_1 = R_2 = R_{\text{iso}}$. Fig.~\ref{fig:Result3} presents the sum-rate performance as a function of transmit power for the considered single antenna user configurations with two antennas at the base stations. The evaluation includes the EIF case as well as joint IRR and EMI scenarios, where the EMI power is fixed at $-65$~dBm. The results clearly show that the presence of joint interference significantly degrades the achievable system sum rate. As the number of UEs increases, the sum rate initially improves but starts to degrade beyond two UEs. This behavior is due to the fixed system resources, such as the number of antennas, transmit power, and RIS size, in this comparison. Moreover, increasing the number of UEs leads to higher inter-UE interference, which further limits the achievable sum rate. For simplicity, each scenario hereafter assumes a setup with a two-antenna BS and two single-antenna users in both clusters, enabling the observation of system behavior in a multi-user environment. It is important to note that the proposed system model and optimization framework are not limited to two users and can be extended to scenarios with a larger number of users and antennas. The two-user setup was selected to maintain clarity of interpretation while clearly illustrating the fundamental performance trends of the proposed approach.

\subsection{Asymptotic Analysis}
\label{Asymptotic_analysis}

\begin{figure}[t!]
    \centering
    \includegraphics[width=0.48\textwidth]{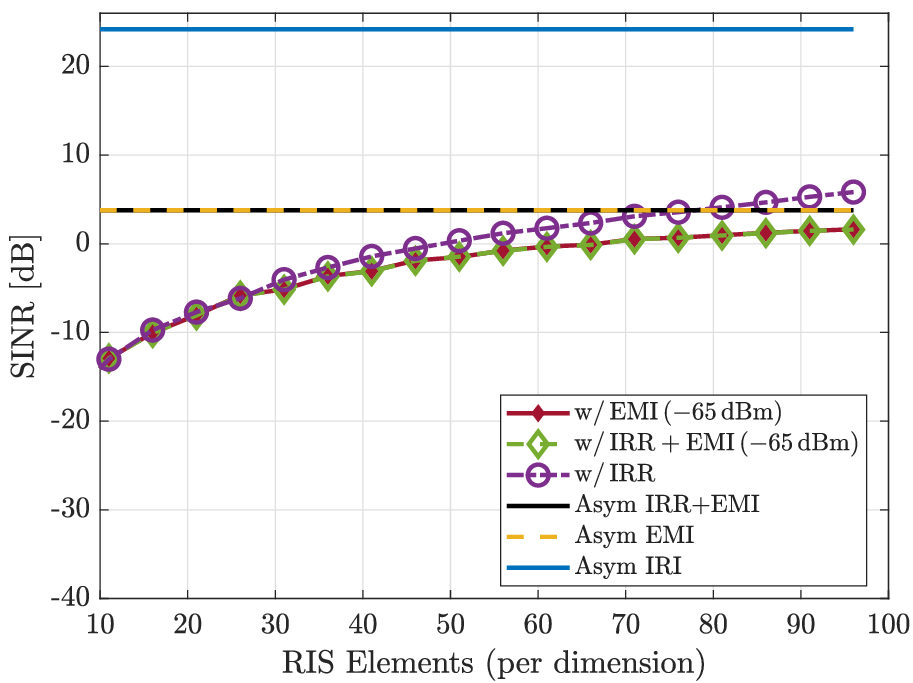}
     %\captionsetup{font=tiny}
    \caption{SINR vs number of RIS elements per dimension under different interference scenarios.}
    \label{fig:Result6}
\end{figure}

\begin{figure}[t!]
    \centering
    \includegraphics[width=\columnwidth]{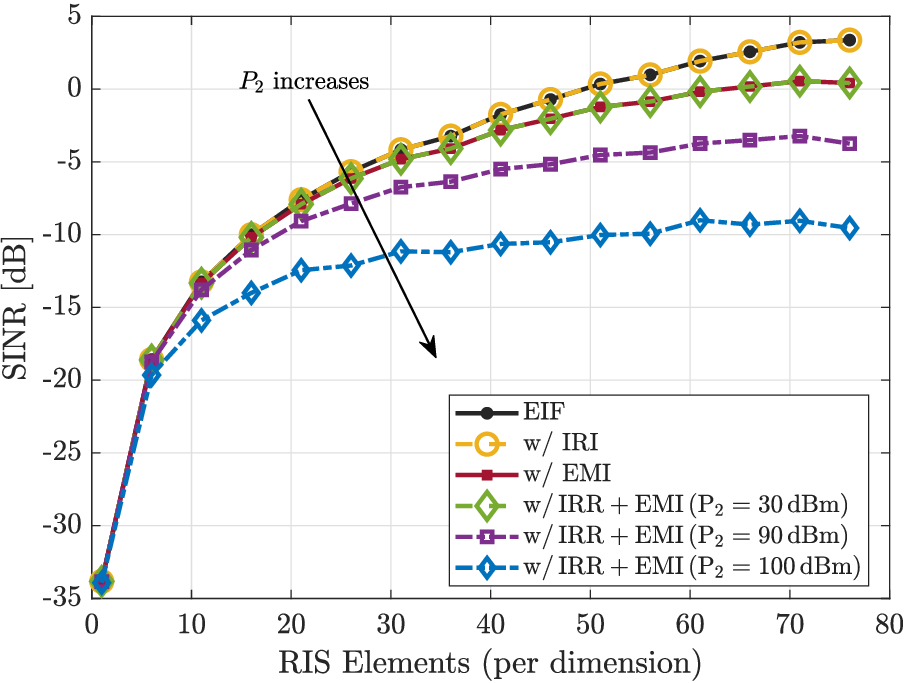}
     %\captionsetup{font=tiny}
    \caption{SINR vs number of RIS elements per dimension for varying interference transmit powers}
    \label{fig:Result5}
%  \vspace{-2em} 
\end{figure}

To gain analytical insight in the asymptotic regime, for tractability and simplicity, we focus 
on the SISO scenario. %, as the MISO case leads to prohibitive analytical complexity. 
Specifically, $\gamma^{\mathrm{EMI+IRR}} \approx \mathcal{N}/\mathcal{D},$ %can be expressed as the ratio $\mathcal{N}/\mathcal{D}$, 
where $\mathcal{N} = p_1 \left| \mathbf{g}_{11}^H 
\boldsymbol{\Theta}_1 \mathbf{h}_1 \right|^2$ represents the desired signal power, and 
$\mathcal{D} = p_2 \left| \mathbf{g}_{11}^H \boldsymbol{\Theta}_1 \mathbf{Z}_{21}^H 
\boldsymbol{\Theta}_2 \mathbf{h}_2 \right|^2 + \mathbf{g}_{11}^H \boldsymbol{\Theta}_1 
A_1\sigma_1^2 \mathbf{R}_1 \boldsymbol{\Theta}_1^H \mathbf{g}_{11} + \mathbf{g}_{11}^H 
\boldsymbol{\Theta}_1 \mathbf{Z}_{21}^H \boldsymbol{\Theta}_2 A_2\sigma_2^2 \mathbf{R}_2 
\boldsymbol{\Theta}_2^H \mathbf{Z}_{21} \boldsymbol{\Theta}_1^H \mathbf{g}_{11}$% + \sigma_w^2$ 
is the interference sources. 
As $N_1 \to \infty$, invoking the law of large numbers,the SINR can be approximated as 
$\gamma^{\mathrm{EMI+IRR}} \approx 
\mathbb{E}[\mathcal{N}]/\mathbb{E}[\mathcal{D}]$. Under the additional assumption 
$\mathbf{R}_n = \mathbf{I}_n$, the individual expectations comprising $\mathbb{E}[\mathcal{N}]$ 
and $\mathbb{E}[\mathcal{D}]$ are evaluated as
$\mathbb{E}\left[ p_{1} \left| \mathbf{g}_{1}^{H} \boldsymbol{\Theta}_{1} \mathbf{h}_{1} \right|^{2} \right] 
= p_{1} A_{1}^{2} \beta_{\mathrm{TX\text{-}RIS}}^{(1)} \beta_{\mathrm{RIS\text{-}RX}}^{(1)} N_1$,
$\mathbb{E}\left[ \mathbf{g}_1^{H} \boldsymbol{\Theta}_1 A_1 \sigma_1^2 \mathbf{R}_1 
\boldsymbol{\Theta}_1^{H} \mathbf{g}_1 \right] = A_1^{2} \sigma_1^2 \beta_{\mathrm{RIS\text{-}RX}}^{(1)} N_1$,
$\mathbb{E}\left[ p_{2} \left| \mathbf{g}_{1}^{H} \boldsymbol{\Theta}_{1} \mathbf{Z}_{21}^{H} 
\boldsymbol{\Theta}_{2} \mathbf{h}_{2} \right|^{2} \right] = p_{2} (A_{1}A_{2})^{3/2} 
\beta_{\mathrm{RIS\text{-}RX}}^{(1)} \beta_{\mathrm{TX\text{-}RIS}}^{(2)} 
\beta_{\mathrm{RIS1\text{-}RIS2}} N_{1} N_{2}$, and
$\mathbb{E}\left[ \mathbf{g}_1^H \boldsymbol{\Theta}_1 \mathbf{Z}_{21}^H \boldsymbol{\Theta}_2 
A_2\sigma_2^2 \mathbf{R}_2 \boldsymbol{\Theta}_2^H \mathbf{Z}_{21} \boldsymbol{\Theta}_1^H 
\mathbf{g}_1 \right] = (A_1 A_2)^{3/2} \sigma_2^2 \beta_{\mathrm{RIS\text{-}RX}}^{(1)} 
\beta_{\mathrm{RIS1\text{-}RIS2}} N_1 N_2$, respectively.
Substituting these expressions into $\mathbb{E}[\mathcal{N}]/\mathbb{E}[\mathcal{D}]$ and 
taking the limit as $N_1 \to \infty$, the asymptotic SINR converges to
\begin{equation}
\lim_{N_1\to\infty}
\gamma_{\mathrm{}}^{\mathrm{EMI+IRR}}
=
\frac{
p_1 A_1^2 \beta_{\mathrm{TX\text{-}RIS}}^{(1)}
}{
\begin{aligned}
&p_2 (A_1A_2)^{3/2} \beta_{\mathrm{TX\text{-}RIS}}^{(2)} \beta_{\mathrm{RIS1\text{-}RIS2}} N_2 +...\\
& (A_1A_2)^{3/2} \sigma_2^2 \beta_{\mathrm{RIS1\text{-}RIS2}} N_2 +...\\
& A_1^2 \sigma_1^2
\end{aligned}
}.\label{Asymptot_EMI-IRR}
\end{equation}

The asymptotic SINR expressions for the individual impairment scenarios can be readily obtained as special cases of \eqref{Asymptot_EMI-IRR}. Specifically, for the IRR-only 
scenario, setting $\sigma_1^2 = \sigma_2^2 = 0$ yields
\begin{equation}
\lim_{N_1\to\infty}
\gamma_{\mathrm{}}^{\mathrm{IRR}}
=
\frac{
p_1 \sqrt{A_1} \beta_{\mathrm{TX\text{-}RIS}}^{(1)}
}{
p_2 A_2^{3/2} \beta_{\mathrm{TX\text{-}RIS}}^{(2)} \beta_{\mathrm{RIS1\text{-}RIS2}} N_2
},\label{Asymptot_IRR}
\end{equation}
whereas for the EMI-only scenario, setting $\beta_{\mathrm{RIS1\text{-}RIS2}} N_2 = 0$ 
gives
\begin{equation}
\lim_{N_1 \to \infty} \gamma_{\mathrm{}}^{\mathrm{EMI}}
=
\frac{
p_1 \beta_{\mathrm{TX\text{-}RIS}}^{(1)}
}{
\sigma_1^{2}
}.\label{Asymptot_EMI}
\end{equation}

As $N_1 \to \infty$, the thermal noise power $\sigma_\text{w}^2$ vanishes entirely, confirming that the system transitions from a noise-limited to a fully interference-limited regime as the RIS scales. The local EMI term $A_1^2\sigma_1^2$ persists as a fixed,
constant in the denominator, independent of $N_1$ and $N_2$, revealing a fundamental performance ceiling that cannot be overcome by simply increasing the number of RIS elements. Furthermore, the IRR-related interference terms in the denominator of \eqref{Asymptot_EMI-IRR} scale linearly with $N_2$, with the inter-RIS signal interference term additionally scaling with $p_2$, implying that increasing either the neighboring RIS size or its transmit power actively degrades performance by amplifying inter-cluster interference through the IRR path, ultimately driving $\gamma^{\mathrm{EMI+IRR}} \to 0$ as $N_2 \to \infty$ or $p_2 \to \infty$. It is also noted that the RIS-to-UE path loss coefficient $\beta_{\mathrm{RIS\text{-}RX}}^{(1)}$ cancels between the numerator and 
denominator, indicating that asymptotic performance is governed by the BS-to-RIS link quality $\beta_{\mathrm{TX\text{-}RIS}}^{(1)}$ and the interference structure rather than the RIS-to-UE distance.

Fig.~\ref{fig:Result6} presents the SINR as a function of the number of elements under all considered interference scenarios. Under severe EMI conditions ($-65$~dBm), a distinct difference emerges. The SINR is significantly reduced across all RIS configurations and converges toward the analytically derived asymptotic ceiling, beyond which increasing the number of RIS elements yields no further performance gain. This saturation behavior reveals a fundamental limitation of RIS-assisted systems when EMI is sufficiently strong, array scaling alone cannot overcome the interference floor. A particularly important observation is that the joint IRR and EMI scenario is virtually indistinguishable from its respective EMI-only counterpart, confirming that IRR contributes negligibly to the overall system degradation when EMI is present, thereby establishing EMI as the dominant performance-limiting impairment. This is analytically confirmed by comparing the asymptotic expressions in \eqref{Asymptot_EMI} and \eqref{Asymptot_EMI-IRR}. 
The additional IRR-related terms appearing in the denominator of \eqref{Asymptot_EMI-IRR} involve cascaded path-loss coefficients across the communication channel 
links, which are severely attenuated in dense indoor factory environments. 
Consequently, these terms become negligible relative to the EMI contribution, and the asymptotic ceiling of the joint scenario 
effectively coincides with that of the EMI-only case providing a closed-form theoretical justification for the simulation observations and underscoring the critical importance of strategic EMI management in RIS-assisted indoor factory deployments.

Fig.~\ref{fig:Result5} further examines the joint impact of EMI and IRR by varying the transmit power of the interfering BS (BS~2). The EIF scenario achieves the highest SINR, growing with the number of RIS elements. The IRR-only scenario is virtually indistinguishable from the EIF curve, further indicating that IRR does not introduce appreciable degradation relative to the interference-free baseline. As the transmit power of BS~2 increases from $30$~dBm to $100$~dBm, a progressive degradation in SINR is observed across all RIS configurations. At extremely high transmit power levels $100$~dBm, however, the sum rate degradation becomes substantially more pronounced, with the gap relative to the interference-free case widening considerably. This result underscores that while IRR may be negligible under typical power conditions, it becomes a significant performance-limiting factor at very high transmit power levels.

%%%%%%%%%%%%%%%%%%%%%%%%%%%%%%%
%%%%%%%%%%%%%%%%%%%%%%%%%%%%%%%
%%%                         %%%
%%%         SECTION         %%%
%%%                         %%%
%%%%%%%%%%%%%%%%%%%%%%%%%%%%%%%
%%%%%%%%%%%%%%%%%%%%%%%%%%%%%%%
\section{Proposed RIS Phase Optimization}
\label{sec:AlternateOptimization}

Due to the presence of EMI and IRR, a significant degradation in system performance can be observed. Also the analytical justification for the simulation results 
presented in Section~\ref{Asymptotic_analysis},  establishes that 
optimization with interference at the RIS controller becomes increasingly essential as the 
deployment scales. To address these impacts, an optimization framework is proposed that jointly optimizes the precoding and RIS phase shifts across three scenarios: EIF, EMI, and joint EMI-IRR. To this end, we consider the sum rate maximization problem for all users, which is formulated in a general mathematical form applicable to all scenarios as
\setcounter{equation}{13}
\begin{equation}\label{eq:optimization_P}
\underset{\boldsymbol{\Theta}_1}{\text{maximize}} \quad 
    f(\boldsymbol{\Theta}_1) 
    = \sum_{k=1}^{K_1}\log(1 + \gamma_{k1}(\boldsymbol{\Theta}_1))
\end{equation}
This unified problem is then optimized and solved individually for each scenario.
\begin{figure*}[t]
\setcounter{equation}{15}

\begin{equation}
\gamma_{k1}(\boldsymbol{\theta}_1)^{\text{EMI+IRR}}  = 
\frac{p_{k1} | \boldsymbol{\theta}_1^\mathrm{H} \mathbf{a}_{k1,k}|^2}
{ \sum_{i \neq k} p_{i1} |\boldsymbol{\theta}_1^\mathrm{H} \mathbf{a}_{k1,i} |^2 
+ \sum_{j \neq k} p_{j1} |\boldsymbol{\theta}_1^\mathrm{H} \mathbf{e}_{k1,j} |^2
+ \boldsymbol{\theta}_1^\mathrm{H} \mathbf{C} \boldsymbol{\theta}_1 + \boldsymbol{\theta}_1^\mathrm{H} \mathbf{D} \boldsymbol{\theta}_1 + \sigma_{\text{w}}^2}.
\label{eq:sinr_expression_new_EMI_IRR}
\end{equation}

\vspace{0.5em} % small vertical space between equations
\setcounter{equation}{18}
\begin{equation}
\begin{split}
\mathbf{c_k} =\ 
& \frac{
2\sum_i p_{i1} \mathbf{a}_{k1,i} \mathbf{a}_{k1,i}^\mathrm{H} \boldsymbol{\theta}_1 + 2\sum_j p_{j2} \mathbf{e}_{k1,j} \mathbf{e}_{k1,j}^\mathrm{H} \boldsymbol{\theta}_1
+(\mathbf{C}+\mathbf{C}^\mathrm{H})\boldsymbol{\theta}_1 + (\mathbf{D}+\mathbf{D}^\mathrm{H})\boldsymbol{\theta}_1 }{
\sum_i p_{i1}| \boldsymbol{\theta}_1^\mathrm{H} \mathbf{a}_{k1,i}|^2 + \sum_j p_{j2}| \boldsymbol{\theta}_1^\mathrm{H} \mathbf{e}_{k1,j} |^2 + \boldsymbol{\theta}_1^\mathrm{H} \mathbf{C} \boldsymbol{\theta}_1 + \boldsymbol{\theta}_1^\mathrm{H} \mathbf{D} \boldsymbol{\theta}_1 + \sigma_{\text{w}}^2
} \\
& \quad \quad \quad \quad \quad \quad \quad  \quad \quad-
\frac{
2\sum_{i \ne k} p_{i1} \mathbf{a}_{k1,i} \mathbf{a}_{k1,i}^\mathrm{H} \boldsymbol{\theta}_1 
+ 2\sum_j p_{j2} \mathbf{e}_{k1,j} \mathbf{e}_{k1,j}^\mathrm{H} \boldsymbol{\theta}_1
+(\mathbf{C}+\mathbf{C}^\mathrm{H})\boldsymbol{\theta}_1 + (\mathbf{D}+\mathbf{D}^\mathrm{H})\boldsymbol{\theta}_1}{
\sum_{i \ne k} p_{i1}| \boldsymbol{\theta}_1^\mathrm{H} \mathbf{a}_{k1,i}|^2 + \sum_{j \ne k} p_{j2}| \boldsymbol{\theta}_1^\mathrm{H} \mathbf{e}_{k1,j} |^2 + \boldsymbol{\theta}_1^\mathrm{H} \mathbf{C} \boldsymbol{\theta}_1 + \boldsymbol{\theta}_1^\mathrm{H} \mathbf{D} \boldsymbol{\theta}_1 + \sigma_{\text{w}}^2
}.
\label{eq:Ck}
\end{split}
\end{equation}
\hrulefill
\end{figure*}
The optimization problem in \eqref{eq:optimization_P} is challenging to solve directly because the objective is non-convex, with SINR terms expressed as ratios of quadratic functions of $\boldsymbol{\Theta}_1$. Moreover, the RIS phase shift variables are constrained to lie on the complex unit circle manifold, which defines a non-convex feasible set. Thus, we adopt an AO framework, where one set of variables is optimized while the other is kept fixed~\cite{guo2020weighted}. AO has been widely used to address similar non-convex problems in the literature. For the considered problem structure, AO can be shown to converge to a stationary point under given conditions~\cite{wu2019intelligent,huang2019reconfigurable,nguyen2022leveraging,guo2020weighted,long2024mmse,ma2023physical}. In our formulation, when the precoding vector $\mathbf{u}_{k1}$ is fixed, the RIS phase shifts $\boldsymbol{\Theta}_1$ are optimized. Conversely, for fixed RIS phases, the precoding vectors are updated in closed form through the zero-forcing (ZF) solution. The proposed phase optimization is investigated under three different scenarios, where in each case the AO framework alternates between optimizing the RIS phases and updating the ZF precoding until convergence. Let us now shift our attention to the phase optimization sub-problem for fixed precoding vector $\mathbf{u}_{k1}$.
\setcounter{equation}{14}

\subsection{\texorpdfstring{Novel Phase Optimization Under Joint EMI and IRR}{Novel Phase Optimization Under Joint EMI and IRR Scenario}} \label{opt_theta_joint_EMI_IRR}

This subsection presents the proposed RIS phase optimization schemes for RIS~1, considering both cases with and without interference information, following the model in Section~\ref{subsec:EMI_IRR_RIS}. Accordingly, the problem in \eqref{eq:optimization_P} can be presented considering $\gamma$ defined as in \eqref{eq:sinr_expression_EMI_IRR}. 
\begin{equation} \label{eq:optimization_newEMI_IRR}
\underset{\boldsymbol{\Theta}_1}{\mathrm{maximize}} \quad 
f_{\text{A}}(\boldsymbol{\Theta}_1) = \sum_{k=1}^{K_1} \log(1 + \gamma_{k1}(\boldsymbol{\Theta}_1)^{\text{EMI+IRR}})
\end{equation}
To make the optimization problem more tractable, we define $\boldsymbol{\theta_1} = [ \theta_{1,1}, \dots, \theta_{1,L_1^2} ]^\mathrm{H}$ and $\mathbf{a}_{k1,i} = \operatorname{diag}(\mathbf{g}_{k1}^{*}) \mathbf{H}_{1} \mathbf{u}_{i1} \in \mathbb{C}^{L_1^2 \times 1}$. Since we consider a quasi-static indoor scenario, and further assuming $\mathbf{C} = \operatorname{diag}(\mathbf{g}_{k1}^{*}) A_1 \sigma_1^2 \mathbf{R}_1 \operatorname{diag}(\mathbf{g}_{k1}) \in \mathbb{C}^{L_1^2 \times L_1^2}$, $\mathbf{D}=\operatorname{diag}(\mathbf{g}_{k1}^{*})\mathbf{Z}_{21}^\mathrm{H} \mathbf{\Theta}_{2}A_2 \sigma_2^2 \mathbf{R}_2\mathbf{\Theta}_{2}^\mathrm{H} \mathbf{Z}_{21}\operatorname{diag}(\mathbf{g}_{k1}) \in \mathbb{C}^{L_1^2 \times L_1^2}$, and $ \mathbf{e}_{k1,j} = \operatorname{diag}(\mathbf{g}_{k1}^{*}) \mathbf{Z}_{21}^\mathrm{H} \mathbf{\Theta}_{2} \mathbf{H}_{21}\mathbf{u}_{j2} \in \mathbb{C}^{L_2^2 \times 1} $, $\gamma_{k1}^{\text{EMI+IRR}}$ can be expressed as in \eqref{eq:sinr_expression_new_EMI_IRR} at the top of the page. Therefore, the optimization problem in \eqref{eq:optimization_newEMI_IRR} takes the form
\setcounter{equation}{16}
\begin{subequations} \label{eq:optimization_newEMI_IRR_tractable}
\begin{align}
\underset{\boldsymbol{\theta}_1}{\mathrm{maximize}} \quad 
& f_{\text{B}}(\boldsymbol{\theta}_1) = \sum_{k=1}^{K_1} \log(1 + \gamma_{k1}(\boldsymbol{\theta}_1)^{\text{EMI+IRR}}) \\
\mathrm{s.t.} \quad 
& |\theta_{1,l}| = 1, \quad \forall l = 1, \dots, L_1^2.
\end{align}
\end{subequations}

It can be observed that \eqref{eq:optimization_newEMI_IRR_tractable} is non-convex. However, the objective function $f_{\text{B}}(\boldsymbol{\theta}_1)$ is continuous and differentiable, and the constraint set defined by $\boldsymbol{\theta}_1$ forms a complex circle manifold. Consequently, a stationary solution to \eqref{eq:optimization_newEMI_IRR_tractable} can be effectively obtained via the RCG algorithm~\cite{guo2020weighted}, which has been widely adopted in RIS-aided communication systems. The employment of the RCG method is motivated by its ability to handle the complex circle manifold on which RIS phase optimization lies, while ensuring convergence to a stationary solution with manageable complexity~\cite{shtaiwi2023sum,misbah2023phase,WilsonManifoldMIMO2024}. During each iteration, the RCG algorithm proceeds through several key steps. First, the Euclidean gradient is computed as
\begin{equation}
\nabla \mathbf{f_{\text{B}}} = \sum_{k=1}^{K_1}\mathbf{c_k},
\label{eq:euclideanGradient_EMI_IRR}
\end{equation}
where $\mathbf{c_k}$ is given in \eqref{eq:Ck} at the bottom of the page. Subsequently, the Riemannian gradient is obtained by projecting the Euclidean gradient \( \nabla \mathbf{f_{\text{B}}} \) onto the tangent space of the complex circle manifold as
\setcounter{equation}{19}
\begin{equation}
\mathrm{rgrad}
\mathbf{f_{\text{B}}} = \nabla \mathbf{f_{\text{B}}} - \Re\{ \nabla \mathbf{f_{\text{B}}} \odot \boldsymbol{\theta_1}^* \} \odot \boldsymbol{\theta_1}.
\label{eq:RCG}
\end{equation}
Then, the search direction can be determined as 
\begin{equation}
\mathbf{d}^{(r)} = \mathrm{rgrad}
\mathbf{f_{\text{B}}} + \tau_1 \mathcal{T}({\mathbf{d}}^{(r-1)}),
\label{eq:direction}
\end{equation}
where $\tau_1$ is the Polak-Ribiere update parameter as~\cite{Absil2008} 
\begin{equation}
\tau_1 = \frac{\mathrm{rgrad} \mathbf{f_{\text{B}}}^\mathrm{H(r)}( \mathrm{rgrad} \mathbf{f_{\text{B}}}^{(r)} - \mathrm{rgrad} \mathbf{f_{\text{B}}}^{(r-1)})}{{\| \mathrm{rgrad} \mathbf{f_{\text{B}}}^{(r-1)}\|}^2}
\label{eq:ploak}
\end{equation}
and $\mathcal{T}(\cdot)$ denotes the vector transport function defined as 
\begin{equation}
\mathcal{T}(\mathbf{d}^{(r)}) = {\mathbf{d}}^{(r-1)} - \Re \{ {\mathbf{d}}^{(r-1)} \odot \boldsymbol{\theta_1}^* \} \odot \boldsymbol{\theta_1},
\label{eq:transport}
\end{equation}
where ${\mathbf{d}}^{(r-1)}$ is the previous search direction. Then, the updated tangent vector is projected back to the complex circle manifold as
\begin{equation}
{\theta_{1,l}} = \frac{({\theta_{1,l}}  + \tau_2 \mathbf{d}^{(r)})}{|({\theta_{1,l}}  + \tau_2 \mathbf{d}^{(r)})|}, \quad \forall l = 1, \dots, L_1^2, 
\label{eq:theta_update}
\end{equation}
where $\tau_2$ is a step size that satisfies the Armijo condition. The complete RCG algorithm for RIS phase optimization incorporating these steps is summarized in Algorithm~\ref{Algorithm1}, with $\Delta_{\text{Inner}} = |\text{Obj}^{(r)} - \text{Obj}^{(r-1)}|$ and where $\text{Obj}^{(r)}$ denotes the sum rate objective value at iteration~$r$. The algorithm iteratively updates the RIS phase vector starting from an initial vector $\boldsymbol{\theta}_1^{(0)}$ while keeping the precoding vector $\mathbf{u}_{k1}$ fixed. The convergence threshold $\epsilon$ determines when the inner loop terminates.

\begin{algorithm}[h]
\caption{RCG Inner Loop for RIS Phase Optimization}
\label{Algorithm1}
\begin{algorithmic}[1]
\State \textbf{Input:} $\mathbf{u}_{k1}$, $\boldsymbol{\theta}_1^{(0)}$, $\epsilon$, $\omega_k$ and parameters required for computing $\gamma_{k1}(\boldsymbol{\theta}_1)^{\text{EMI+IRR}}$
\State \textbf{Output:} Optimized RIS phase vector $\boldsymbol{\theta}_1^{\text{opt}}$
\State Initialize $r = 0$ and $\mathbf{d} = 0$
\Repeat
    \State Compute the Euclidean gradient using \eqref{eq:sinr_expression_new_EMI_IRR} and \eqref{eq:Ck}
    \State Compute the Riemannian gradient using \eqref{eq:RCG}
    \If{$r = 0$}
        \State $\mathbf{d} = \mathrm{rgrad} \mathbf{f_{\text{B}}}$
    \Else
        \State Compute the update parameter $\tau_1$ as in \eqref{eq:ploak}
        \State Update search direction $\mathbf{d}$ using \eqref{eq:direction}
    \EndIf
    \State Perform Armijo line search to determine step size $\tau_2$
    \State Update $\boldsymbol{\theta_1}^{(r)}$ using \eqref{eq:theta_update}
    \State Evaluate objective: $\text{Obj}^{(r)} \gets f(\boldsymbol{\theta_1}^{(r)})$
    \State Store $\mathrm{rgrad} \mathbf{f_{\text{B}}}^{(r-1)} \gets \mathrm{rgrad} \mathbf{f_{\text{B}}}^{(r)}$, $\mathbf{d}^{(r-1)} \gets \mathbf{d}^{(r)}$
    \State $r = r + 1$
\Until{$|\Delta_{\textbf{Inner}}|  \leq \epsilon$ }
\State \textbf{Return:} $\boldsymbol{\theta_1}^{\text{opt}} \gets \boldsymbol{\theta_1}^{(r)}$
\end{algorithmic}
\end{algorithm}

\subsection{\texorpdfstring{Phase Optimization Under EMI and EIF}{Phase Optimization under EMI and EIF}}
\label{Opt_EMI_EIF}

This section presents the baseline optimization formulations, which serve as reference benchmarks for the proposed joint EMI-IRR optimization framework and enable a comprehensive comparison between the joint and individual interference cases. We now assume that cluster~1 is in the presence and absence of EMI as described in section~\ref{subsec:EMI_RIS} and ~\ref{subsec:IF} respectively. In line with the previous formulation, we retain the same definition of $\boldsymbol{\theta}_1$. In the absence of cluster~2, the IRR term vanishes, which leads to $\mathbf{D} = \mathbf{0}$ and $\mathbf{e}_{k1,j} = \mathbf{0}$ in \eqref{eq:sinr_expression_new_EMI_IRR}. Consequently, $\gamma_{k1}^{\text{EMI}}$ in \eqref{eq:sinr_expression_EMI} can be written as
\begin{equation} \label{eq:sinr_expression_new_EMI}
\gamma_{k1}(\boldsymbol{\theta}_1)^{\text{EMI}} = \frac{p_{k1} | \boldsymbol{\theta}_1^\mathrm{H} \mathbf{a}_{k1,k} |^2}
{\sum_{i \neq k} p_{i1} |\boldsymbol{\theta}_1^\mathrm{H} \mathbf{a}_{k1,i} |^2 
+ \boldsymbol{\theta}_1^\mathrm{H} \mathbf{C}\boldsymbol{\theta}_1 + \sigma_{\text{w}}^2}.
\end{equation}

Similarly, by setting $C =0$ in \eqref{eq:sinr_expression_new_EMI}, $\gamma_{k1}(\boldsymbol{\theta}_1)^{\text{EIF}}$ can be expressed as 
\begin{equation} \label{eq:sinr_expression_newIF}
\gamma_{k1}(\boldsymbol{\theta}_1)^{\text{EIF}} = \frac{p_{k1} | \boldsymbol{\theta}_1^\mathrm{H} \mathbf{a}_{k1,k} |^2}
{\sum_{i \neq k} p_{i1} |\boldsymbol{\theta}_1^\mathrm{H} \mathbf{a}_{k1,i} |^2 + \sigma_{\text{w}}^2}.
\end{equation}

Considering the original optimization problem in \eqref{eq:optimization_newEMI_IRR_tractable}, the same RCG-based framework described earlier can be applied to obtain a stationary solution. Specifically, by modifying the objective function using its Euclidean gradient, given by

\begin{equation}
\nabla \mathbf{f} = \sum_{k=1}^{K_1} \alpha,
\label{eq:alpha}
\end{equation}

where, 

\begin{equation}
\boldsymbol{\alpha}_k =
\begin{cases}
\mathbf{b}_k, & \gamma =\gamma_{k1}(\boldsymbol{\theta}_1)^{\text{EMI}}, \\
\mathbf{a}_k, & \gamma =\gamma_{k1}(\boldsymbol{\theta}_1)^{\text{EIF}}.
\end{cases}
\label{eq:alpha_k}
\end{equation}

The term $\mathbf{b}_k$ is given by

\begin{equation}
\begin{split}
\mathbf{b_k} =\ 
&\frac{
2\sum_i p_{i1} \mathbf{a}_{k1,i} \mathbf{a}_{k1,i}^\mathrm{H} \boldsymbol{\theta}_1 
+(\mathbf{C}+\mathbf{C}^\mathrm{H})\boldsymbol{\theta}_1}{
\sum_i p_{i1}| \boldsymbol{\theta}_1^\mathrm{H} \mathbf{a}_{k1,i} |^2 + \boldsymbol{\theta}_1^\mathrm{H} \mathbf{C} \boldsymbol{\theta}_1 + \sigma_{\text{w}}^2
} \\
& \quad \quad \quad-
\frac{
2\sum_{i \ne k} p_{i1}\mathbf{a}_{k1,i} \mathbf{a}_{k1,i}^\mathrm{H} \boldsymbol{\theta}_1
+(\mathbf{C}+\mathbf{C}^\mathrm{H})\boldsymbol{\theta}_1}{
\sum_{i \ne k} p_{i1}| \boldsymbol{\theta}_1^\mathrm{H} \mathbf{a}_{k1,i} |^2 + \boldsymbol{\theta}_1^\mathrm{H} \mathbf{C} \boldsymbol{\theta}_1 + \sigma_{\text{w}}^2
}.
\end{split}
\label{eq:Bk}
\end{equation}

Similarly, $\mathbf{a}_k$ is expressed as

\begin{equation}
\begin{split}
\mathbf{a_k} =\ 
&\frac{
\sum_i p_{i1} \mathbf{a}_{k1,i} \mathbf{a}_{k1,i}^\mathrm{H} \boldsymbol{\theta}_1 
}{
\sum_i p_{i1} | \boldsymbol{\theta}_1^\mathrm{H} \mathbf{a}_{k1,i} |^2 + \sigma_{\text{w}}^2
} -
\frac{
\sum_{i \ne k} p_{i1} \mathbf{a}_{k1,i} \mathbf{a}_{k1,i}^\mathrm{H} \boldsymbol{\theta_1} 
}{
\sum_{i \ne k} p_{i1} | \boldsymbol{\theta_1}^\mathrm{H} \mathbf{a}_{k1,i} |^2 + \sigma_{\text{w}}^2
}.
\label{eq:Ak}
\end{split}
\end{equation}

\subsection{Precoder Optimization} \label{PrecoderOpt}

In this work, we adopt ZF precoding given by $\mathbf{H}_{\text{eff}}^\mathrm{H} ( \mathbf{H}_{\text{eff}} \mathbf{H}_{\text{eff}}^\mathrm{H} )^{-1}$ where the effective channel is given by \(\mathbf{H}_{\text{eff}} = \mathbf{G}_{1} \boldsymbol{\Theta}_1 \mathbf{H}_1\).  Here, $\mathbf{G}_{1} = [\mathbf{g_{11}^\mathrm{H}}, \ldots, \mathbf{g_{k1}^\mathrm{H}}]^T$. This approach is applied across all clusters and scenarios for consistency and simplicity of implementation and analysis. When $\boldsymbol{\Theta}_1$ is optimized via Algorithm~\ref{Algorithm1}, 
the resulting phase shift values can be directly substituted into the closed-form precoder expressions to obtain the corresponding precoder outputs. In our AO framework, we employ the ZF scheme to compute $\mathbf{u}_{k1}$ for better trackability. The outer loop of the AO is summarized in Algorithm~\ref{Algorithm2}, where $\eta$ is the convergence threshold, $\Delta_{\textbf{Outer}} = | \text{Obj}^{(t)} - \text{Obj}^{(t-1)} |$.

\begin{algorithm}[h]
\caption{AO Outer Loop for Joint RIS and Precoding}
\label{Algorithm2}
\begin{algorithmic}[1]
\State \textbf{Input:} $\boldsymbol{\theta}_1^{(0)}$, $\eta$, $\omega_k$ and parameters required for computing $\mathbf{u}_{k1}$, $\gamma_{k1}(\boldsymbol{\theta}_1)^{\text{EMI+IRR}}$
\State \textbf{Output:} Optimized RIS matrix $\boldsymbol{\Theta}_1^{\text{opt}}$
\State Initialize $t = 0$ and $\boldsymbol{\theta}_1 = \boldsymbol{\theta}_1^{(0)}$
\Repeat 
    \State Compute $\mathbf{u}_{k1}$ using ZF precoder 
    \State For fixed $\mathbf{u}_{k1}$, optimize $\boldsymbol{\theta}_1$ using Algorithm~\ref{Algorithm1}
    \State Store: $\text{Obj}^{(t)} \gets$ final objective value from Algorithm~\ref{Algorithm1}
    \State $t = t + 1$
\Until{$\Delta_{\textbf{Outer}} \leq \eta$}
\State \textbf{Return:} $\boldsymbol{\Theta}_1^{\text{opt}} = \mathrm{diag}(\boldsymbol{\theta}_1)$
\end{algorithmic}
\end{algorithm}

\subsection{Complexity Analysis}
\label{sec:ComplexityAnalysis}

In the proposed AO framework, the optimization alternates between updating the RIS phase vector and the BS precoder. Under the adopted ZF strategy, the precoder update involves constructing the effective cascaded channel matrix and performing a matrix inversion, which incur computational costs of $\mathcal{O}(K_1^{2}T_1)$ and $\mathcal{O}(K_1^{3})$, respectively. For the RIS phase optimization, the RCG algorithm is employed. The most computationally demanding operation in this procedure is the Euclidean gradient evaluation, which scales as $\mathcal{O}(K_1^{2}L_1^{4})$. Other steps, such as the Armijo backtracking line search and the retraction operation, require only $\mathcal{O}(K_1^{2}L_1^2)$ and $\mathcal{O}(L_1^2)$ computations, respectively, and thus become negligible when $L_1^2$ is large. By combining these results, the overall computational complexity of the AO algorithm can be expressed as $\mathcal{O}\big(I_O (K_1^{2}T_1 + K_1^{3} + I_R K_1^{2}L_1^{4})\big),$
where $I_O$ and $I_R$ denote the number of outer AO iterations and inner RCG iterations, respectively. Overall computational complexities calculated for two sets of parameter settings are provided in Table~\ref{tab:complexity}, assuming $I_O = 10$ and $I_R = 20$.

\begin{table}[t]
\caption{Computational complexity for different parameter settings.}
\centering
\footnotesize
\begin{tabular}{c c c c p{0.42\columnwidth}}
\hline
\textbf{Setting} & $K_1$ & $L_1$ & $T_1$ & \textbf{Overall Complexity} \\
\hline
1 & 2 & 10 & 2 &
$\mathcal{O}\!\big(10(2^{2}\!\times\!2 + 2^{3} + 20\!\times\!2^{2}\!\times\!10^{4})\big)$ \\
& & & &
$\approx \mathcal{O}(8\times10^{6})$ \\
\hline
2 & 4 & 12 & 4 &
$\mathcal{O}\!\big(10(4^{2}\!\times\!4 + 4^{3} + 20\!\times\!4^{2}\!\times\!12^{4})\big)$ \\
& & & &
$\approx \mathcal{O}(6.6\times10^{7})$ \\
\hline
\end{tabular}
\label{tab:complexity}
\end{table}

Here, convergence is achieved based on a threshold-based termination criterion. This convergence threshold enables a controlled trade-off between accuracy and computational cost, thereby supporting feasibility in real-time and large-scale scenarios.

%%%%%%%%%%%%%%%%%%%%%%%%%%%%%%%
%%%%%%%%%%%%%%%%%%%%%%%%%%%%%%%
%%%                         %%%
%%%         SECTION         %%%
%%%                         %%%
%%%%%%%%%%%%%%%%%%%%%%%%%%%%%%%
%%%%%%%%%%%%%%%%%%%%%%%%%%%%%%%
\section{Performance Evaluation with Optimized Phase Shifts}
\label{sec:PerfEvalOpt}

%\subsection{Analysis of EIF, EMI and IRR under Optimized Phase Shifts}

In this section, we investigate the optimization of RIS phase shifts under various interference scenarios. Since the RIS demonstrates resilience to low-level interferences, the phase optimization specifically targeting IRR is omitted as its performance closely overlaps with the EIF scenario. Therefore, our analysis focuses on optimizing RIS phase shifts of cluster~1 with and without the interference information of EMI and IRR at RIS, and to examine their respective behaviors. For all the scenarios discussed, the phase shifts of the RIS in cluster~2 are optimized without the knowledge of interference information at RIS under the EIF scenario presented in Section~\ref{Opt_EMI_EIF}. Hereafter, the terms `interference-aware' and `interference-unaware', or equivalently `with' and `without interference knowledge', refer to the cases where interference information at RIS~1 is available and unavailable, respectively.

Furthermore, an EMI threshold of $-60$~dBm is adopted in the analysis related to phase shift optimization. To simplify the analysis, we assume that the power is equally allocated among the users, with each user receiving a unit power allocation, and each scenario assumes a setup with two antenna BS and two single antenna users, in both clusters.
%\begin{figure}[t!]
%    \centering
%    \includegraphics[scale=0.265]{Result7.eps}
%    \caption{Sum rate vs Transmit power with optimized phase and fixed phase scenarios under MMSE precoding.}
%    \label{fig:Result7}
%\end{figure}
%
%
%\begin{figure}[t!]
%    \centering
%    \includegraphics[scale=0.265]{SumRatesOptFinalPlot.eps}
%    \caption{Sum rate vs Transmit power with optimized phase with and without interference mitigation under MMSE precoding.}
%    \label{fig:Result8}
%\end{figure}
Initially, the transmit power of cluster~1 is varied, considering an RIS with 225 elements in cluster~1. The corresponding phase shifts are optimized without the knowledge of interference information at RIS under the EIF scenario presented in Section~\ref{Opt_EMI_EIF}. All other system parameters are as described in Table~\ref{tab:parameters}. 

\begin{figure}[!t]
    \centering
    \includegraphics[width=\columnwidth]{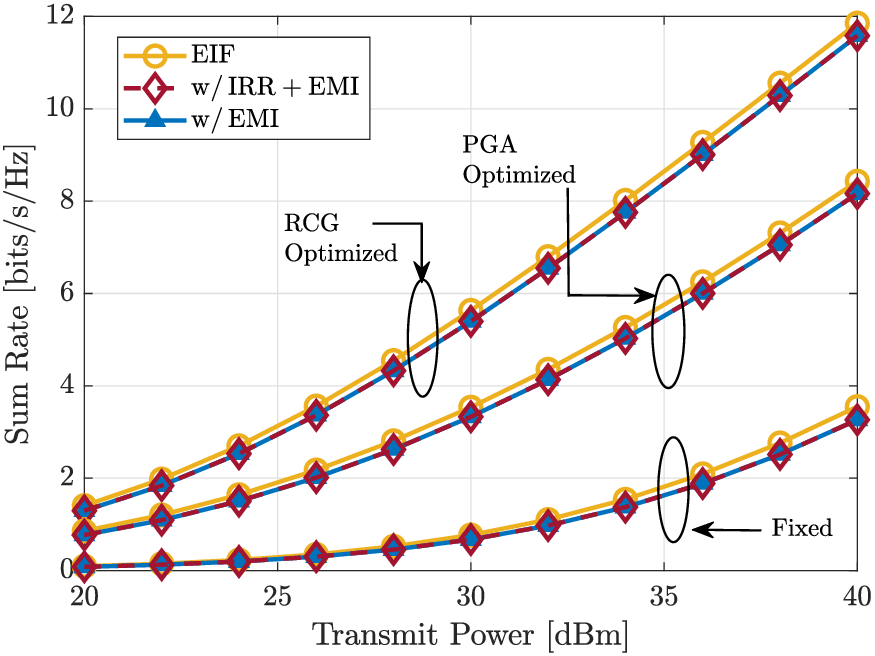}
     %\captionsetup{font=tiny}
    \caption{Sum rate vs transmit power with optimized phase and fixed phase scenarios.}
    \label{fig:Result7}
\end{figure}

\begin{figure}[!t]
    \centering
    \includegraphics[width=\columnwidth]{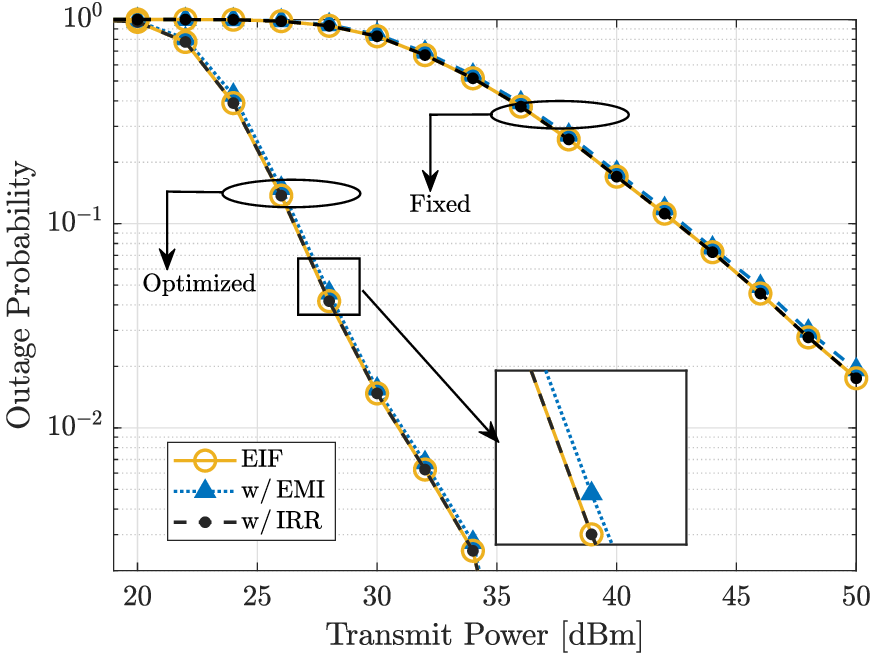}
     %\captionsetup{font=tiny}
    \caption{Outage probability of user 1 vs transmit power under optimized phase and fixed phase scenarios.}
    \label{fig:Result4}
%  \vspace{-2em} 
\end{figure}

%\begin{figure}[t!]
%    \centering
%    \includegraphics[width=\columnwidth]{Result8.eps}
%     %\captionsetup{font=tiny}
%    \caption{Sum rate vs transmit power under optimized phase\\
%    with and without interference knowledge.}
%    \label{fig:Result8}
%\end{figure}

\begin{figure}[t!]
    \centering
    \includegraphics[width=\columnwidth]{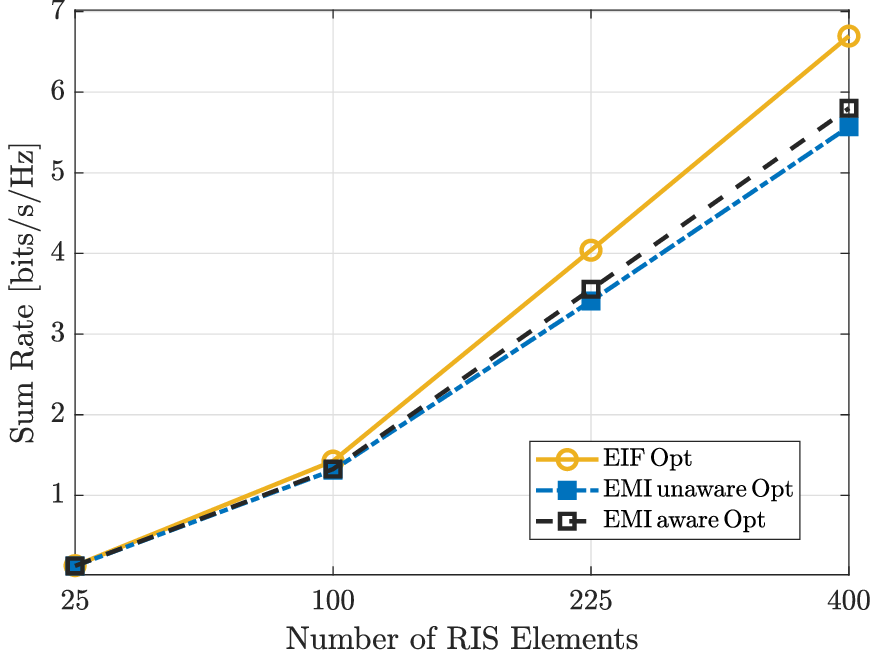}
     %\captionsetup{font=tiny}
    \caption{Sum rate vs number of RIS elements under optimized phase with and without interference knowledge.}
    \label{fig:ElementsVsSumRate}
%  \vspace{-2em} 
\end{figure}

\begin{figure}[t!]
    \centering
    \includegraphics[width=\columnwidth]{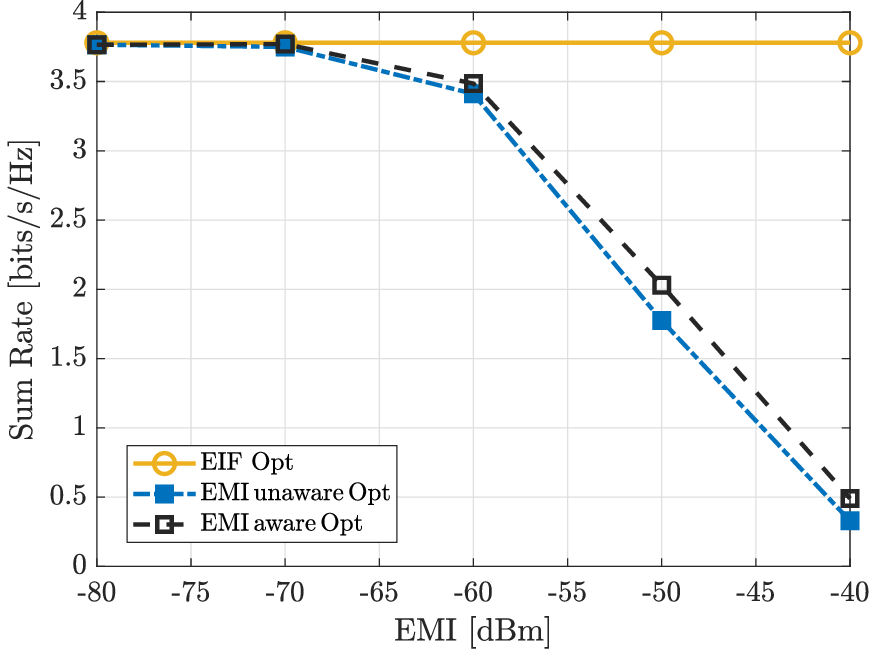}
    \caption{Sum rate vs EMI power under optimized phase with and without interference knowledge.}
    \label{fig:sRateVsEMI}
\end{figure}

The resulting performance is illustrated in Fig.~\ref{fig:Result7}, depicting the sum rate as a function of the transmit power for different interference scenarios with and without phase optimization. As a baseline comparison, we adopted the PGA method, presented in \cite{ArthurBaseline}, which is a widely used optimization approach for RIS phase design. The results demonstrate that the RCG-based optimization framework consistently achieves better performance than the PGA method, highlighting the robustness and effectiveness of the proposed RCG framework. Furthermore, the results show a significant improvement of up to $100\%$ in the achievable sum rate with the RCG-based optimized phase shifts across all interference scenarios compared to the fixed-phase configuration. The phase-optimized results further confirm that IRR remains a negligible impairment under typical operating power levels, with EMI remaining the dominant interference source. In light of this, the EMI-only scenario (only cluster~1 is present) is adopted hereafter as the primary interference model, both for analytical tractability and to maintain clarity in the presentation of subsequent results.

Fig.~\ref{fig:Result4} shows the outage probability of user~1 as a function of the transmit power, with and without RIS phase optimization. Here, the $R_{\text{TH}}$ is $0.1$~bits/s/Hz. With increasing transmit power, the outage probability decreases for both optimized and fixed RIS configurations. However, the optimized RIS achieves the same outage levels at significantly lower transmit power, and the steeper slope of its curves indicates that RIS optimization leads to a much higher rate of improvement in the outage probability. Optimizing the RIS phase shifts constructively aligns the BS to RIS signal with the signals reflected by the RIS. This improvement increases the overall SINR, thereby enhancing the sum rate and subsequently reducing the outage probability.

Next, we optimize the RIS phase shifts in cluster~1 by incorporating interference knowledge of EMI, as described in Section~\ref{Opt_EMI_EIF}. Fig.~\ref{fig:ElementsVsSumRate} illustrates the variation of the sum rate with the number of RIS elements in cluster~1, whereas Fig.~\ref{fig:sRateVsEMI} depicts the impact of different EMI power levels on the sum rate. As shown in Fig.~\ref{fig:ElementsVsSumRate}, the sum rate increases with the number of RIS elements since a larger number of elements provides more variables to optimize over and enables more reflections that constructively enhance the signal strength and improve the SINR. Since the RIS passively reflects all incident signals, including interferences, both desired and interfering signals experience higher reflection gain as the number of elements increases. Consequently, when EMI is present, it also reflected and amplified, causing its impact to intensify with an increasing number of RIS elements. At the same time, the performance gap between the EIF case and the interference-affected cases widens with the increasing number of RIS elements. This occurs because the optimized phase shifts that enhance the desired signal also amplify the interference components, further widening the difference in achievable sum rates.

Fig.~\ref{fig:sRateVsEMI} indicates a decline in sum rate as EMI power increases. For this scenario, an RIS with $225$ elements is used in cluster~1. As EMI gets stronger, the presence of interference knowledge allows the optimization to adjust phase shifts more effectively. As a result, the difference between interference-aware and interference-unaware optimization becomes larger. In both cases, interference-aware optimization consistently outperforms its interference-unaware counterpart. Furthermore, the advantage of interference awareness becomes increasingly significant as the number of RIS elements grows or as the EMI power intensifies. This is because, when interference knowledge is available, the phase optimization process can adaptively address interference effects. A larger RIS provides more controllable elements for phase adjustments when interference knowledge is present. As EMI gets stronger, the availability of interference knowledge enables the RIS to adapt its phase configuration more effectively to address dominant interference. The results confirm that as interference levels increase, adaptive phase shift optimization becomes increasingly crucial to maintain system efficiency. Hence, the proposed AO-based optimization framework demonstrates its capability to address interference and enhance the overall sum rate performance under varying environmental conditions. 

\begin{figure}[t!]
    \centering
    \includegraphics[width=\columnwidth]{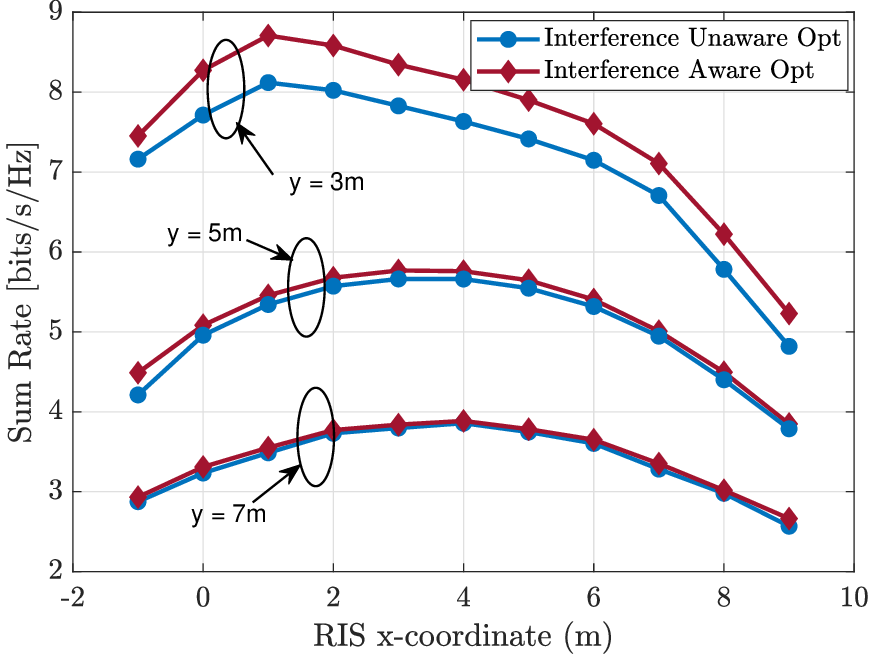}
    \caption{Sum rate vs  RIS deployment location under optimized phase with and without interference knowledge.}
    \label{fig:loaction}
\end{figure}

In Fig.~\ref{fig:loaction}, the $x$-coordinate of RIS~1 is varied while considering three different $y$-coordinate settings, namely $y = 3,\text{m}$, $y = 5,\text{m}$, and $y = 7,\text{m}$. The corresponding system sum rate is evaluated as a function of the RIS deployment location. It can be observed that as the vertical distance of RIS~1 increases, the sum rate decreases due to the increased Euclidean distances between the BS--RIS and RIS--UE links. For a fixed $y$-coordinate, the sum rate initially increases with the $x$-coordinate and then gradually decreases, indicating the presence of an optimal RIS deployment location for each scenario. This optimal location depends on the $y$-coordinate and reflects a trade-off between maximizing the desired signal strength and limiting the impact of interference. As observed from the results, decreasing the vertical distance of RIS~1 from $7\,\text{m}$ to $3\,\text{m}$ shortens the BS--RIS and RIS--UE link distances, which improves the achievable sum rate due to stronger signal strength. Sum rate slightly peaks when the RIS is closer to BS and UEs. However, the performance gap between the interference-aware and interference-unaware optimization frameworks also becomes more pronounced.%, since the impact of IRR increases as the inter-RIS distance decreases. 
This experiment further justifies the effectiveness of interference-aware optimization when interference effects become more significant.

It is important to note that the presented analysis relies on certain modeling assumptions regarding interferences. In practical indoor factory deployments, EMI may originate from various industrial equipment and can exhibit different spatial and temporal characteristics than those considered in the simplified interference model adopted in this study. In addition, the dimension, number, and positioning of RIS panels may vary depending on the factory layout, environmental conditions, and coverage requirements. These assumptions are chosen to avoid unnecessarily complicating the analysis and enable extracting meaningful insights. The proposed optimization framework provides valuable insights into how interference-aware RIS phase optimization can mitigate performance degradation in realistic scenarios. In practical deployments, the algorithm can be adapted by incorporating environment-specific interference measurements or channel estimation techniques to more accurately capture the characteristics of EMI and IRR.

%%%%%%%%%%%%%%%%%%%%%%%%%%%%%%%
%%%%%%%%%%%%%%%%%%%%%%%%%%%%%%%
%%%                         %%%
%%%         SECTION         %%%
%%%                         %%%
%%%%%%%%%%%%%%%%%%%%%%%%%%%%%%%
%%%%%%%%%%%%%%%%%%%%%%%%%%%%%%%
\section{Conclusion}
\label{sec:conc}
In this study, we have investigated the practical challenge posed by the joint impact of EMI and IRR on a downlink multi-RIS-aided MISO communication system in local 6G networks, such as an indoor factory environment. Comprehensive system-level simulations demonstrate that ignoring the impact of EMI and IRR, as done in most studies, leads to over-optimistic performance gains. We have shown that in dense, cluttered environments, across typical interference power levels, system degradation is driven almost entirely by EMI, with IRR exerting a negligible influence. However, at very high transmit powers, IRR becomes a significant impairment that cannot be neglected.

Furthermore, an interference-aware AO-based RIS phase shift optimization algorithm is proposed to overcome the performance loss. The proposed AO framework, integrated with the RCG method, provides a clearer understanding of how interference awareness influences the performance of RIS phase shift optimization. The results further demonstrated that the proposed AO approach can effectively mitigate interferences to some extent. As a future research direction, this study could be extended to scenarios with multiple RISs and to investigate the behavior of EMI and IRR in the context of beyond-diagonal RIS architectures.

\bibliographystyle{IEEEtran}
\bibliography{references}

\end{document}